\documentclass[]{openjournal}

\usepackage{latexsym}
\usepackage{graphicx}
\usepackage{amssymb}
\usepackage{longtable}
\usepackage{epsf}
\usepackage{amsmath}
\usepackage{graphicx}
\graphicspath{ {figures/}}
\usepackage{gensymb}
\DeclareMathOperator{\sech}{sech}
\usepackage{hyperref}
\usepackage{cleveref}
\usepackage[title]{appendix}

\shorttitle{Disk vertical structure}
\shortauthors{Warren S. J., Ahmed S., and Laithwaite R.C.}

\begin{document}

\title{The local vertical density distribution of ultracool dwarfs M7 to L2.5 and their luminosity function}

\author{S.J.~Warren$ ^1$, S. Ahmed$ ^2$, and R.C.~Laithwaite$^1$}
\affil{$^1$Astrophysics Group, Imperial College London, Blackett
  Laboratory, Prince Consort Road, London SW7 2AZ, UK}
\affil{$^2$Centre for Electronic Imaging, School of Physical Sciences, The Open University, Walton Hall, Milton Keynes, MK7 6AA, UK}

\maketitle

We investigate the form of the local vertical density profile of the stars in the Galactic disk, close to the Galactic plane. We use a homogeneous sample of 34\,000 ultracool dwarfs M7 to L2.5 that all lie within 350\,pc of the plane. We fit a profile of the form $\sech^\alpha$, where $\alpha=2$ is the theoretically preferred isothermal profile and $\alpha=0$ is the exponential function. Larger values of $\alpha$ correspond to greater flattening of the profile towards the plane. We employ a likelihood analysis that accounts in a direct way for unresolved binaries in the sample, as well as for the spread in absolute magnitude $M_J$ within each spectral sub-type (Malmquist bias). We measure $\alpha=0.29^{+0.12}_{-0.13}$. The $\alpha=1$ ($\sech$) and flatter profiles are ruled out at high confidence for this sample, while $\alpha=0$ (exponential) is included in the $95\%$ credible interval. Any flattening relative to exponential is modest, and is confined to within 50\,pc of the plane. The measured value of $\alpha$ is consistent with the results of the recent analysis by Xiang et al. Our value for $\alpha$ is also similar to that determined for nearby spiral galaxies by de Grijs et al., measured from photometry of galaxies viewed edge on. The measured profile allows an accurate determination of the local space density of ultracool dwarfs M7 to L2.5, and we use this to make a new determination of the luminosity function at the bottom of the main sequence. Our results for the luminosity function are a factor two to three lower than the recent measurement by Bardalez Gagliuffi et al., that uses stars in the local 25\,pc radius bubble, but agree well with the older study by Cruz et al.\\


\vspace{1 cm}

\twocolumngrid

\section{Introduction}\label{sec:intro}

The variation of the space density of stars in the disk of the Milky Way, in the vertical direction, i.e. perpendicular to the plane of the disk, and at the solar radius, approximates to an exponential distribution \citep{1GilmoreReid} up to heights of 1\,kpc. What happens close to the plane? Is there a sharp density peak, or does the exponential soften? We do not have a clear answer to this question for the Milky Way, but the density profile is often modelled by a $\sech^2$ distribution \citep[e.g.][]{Gould1996,1Siegel,1Ferguson,Bennett2019}, which softens by a factor four relative to an exponential. A self-gravitating isothermal sheet has this equilibrium solution \citep{0Spitzer,Camm1950,vdk81}, and this may be why the $\sech^2$ distribution is popular, even though it is well known that the velocity dispersion of the stars in the disk depends on age.

In considering this question a useful flexible functional form for the density distribution as a function of height $z$ from the plane, is the generalised $\sech$ distribution proposed by \citet{Kruit1988}:
\begin{equation} 
\rho(z)=2^{-2/n}\rho_e\sech^{2/n}(nz/2z_e)\:,
\label{eq:vdk}
\end{equation} 
where $z_e$ is a scale height. With this parameterisation, the exponential, $\sech$, and $\sech^2$ distributions correspond to $n=\infty, 2,$ and 1 respectively.

For data analysis this representation is unsatisfactory, because we want to constrain the value of the parameter $n$, but it has an infinite range, causing difficulty in defining the prior. For this reason we prefer the form used by \citet{Dobbie}, who substitute $\alpha=2/n$, so the function becomes:
\begin{equation}
\rho(z)=2^{-\alpha}\rho_e\sech^{\alpha}(z/\alpha z_e)\:,
\label{eq:dob}
\end{equation} 
and now the exponential, $\sech$, and $\sech^2$ distributions correspond to $\alpha=0,1,2$.

\begin{figure}
\centering
\includegraphics[width = 9.cm, trim = 1cm 8cm 0cm 4cm]{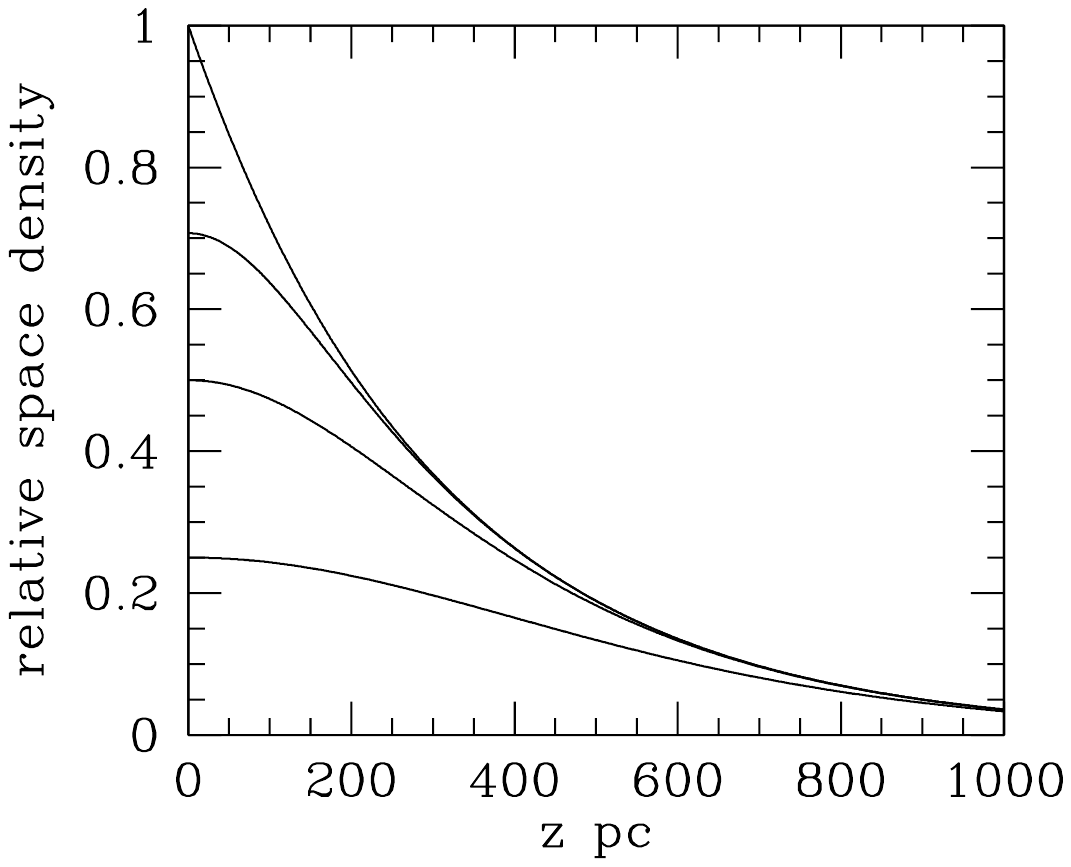}
\caption{Density profiles, according to equation \ref{eq:dob}, with $\rho_e=1.0$ and $z_e=300$\,pc. From top to bottom, the curves are: exponential, $\sech^{0.5}$, $\sech$, $\sech^2$, which are $\alpha=0,0.5,1,2$ respectively.}
\label{fig:sechalpha}
\end{figure}

In this form, with different values of $\alpha$, the functions all have the same density at large values of $|z|$ where they each asymptote to the exponential distribution with scale height $z_e$.  The term $2^{-\alpha}$ is therefore the degree of softening in the centre relative to the exponential distribution. This shows that the $\sech^2$ distribution softens by a factor four, as quoted above. Example functions, with $\rho_e=1$, are plotted in Fig. \ref{fig:sechalpha} for values of $\alpha=0,0.5,1,2,$ (top to bottom, respectively) and for a scale height $z_e=300$\,pc, which is the canonical value for the Milky Way \citep[e.g.][]{1GilmoreReid,1Juric,1Bochanski,1Chang}. 

Beyond a few scale heights there is an excess in the tail, requiring a second population, of larger scale height. However in the current paper we are only concerned with the density distribution close to the plane, at heights $|z|<350$\,pc, so we will assume a single population.

In the remainder of this paper we will use the following equivalent parameterisation of the density function:
\begin{equation}
\rho(z)=\rho_0\sech^{\alpha}(z/\alpha z_e)=\rho_0\sech^{\alpha}((z^{\prime}+z_\odot)/\alpha z_e)\:.
\label{eq:war}
\end{equation} 
Here $\rho_0$ is the density in the Galactic plane, $z_\odot$ is the height of the Sun above the Galactic plane, and $z^{\prime}=z-z_\odot$ is the vertical height measured from the Sun. Because our analysis is confined to distances $<350\,$pc we ignore any effect of variation of the stellar density with Galactic radius, because the scale length for this is so large, $\sim3$\,kpc.

\citet{Dobbie} summarise the status of measurements of $\alpha$ in external galaxies and in the Milky Way. The best study of $\alpha$ in external galaxies is the analysis by \citet{deGrijs} who measured the surface brightness profiles of edge-on spiral galaxies in the $K$ band, to minimise the  effects of extinction. From a sample of 24 galaxies they found a distribution of values of $\alpha=0.5\pm0.2$ (corrected for seeing and extinction), and they argue that the true value may be even lower due to a bias because the galaxies are not viewed perfectly edge on.

In the Milky Way itself there have been very few quantitative studies that contribute to this question. Using observations in the near-infrared \citet{Hammersley1999} state that ``Analysis of one relatively isolated cut through an arm near longitude 65 degrees categorically precludes any possibility of a $\sech^2$ stellar density function for the disc.'' In a footnote \citet{1Juric} state that the exponential profile provides a better fit than $\sech^2$ close to the plane. Using {\em Gaia DR1} \citet{Bovy2017} draws a different conclusion. He measured the vertical density distribution separately for different spectral types A to K, and states ``All vertical profiles are well represented by $\sech^2$ profiles, with scale heights ranging from $\sim50$\,pc for A stars to $\sim150$\,pc for G and K dwarfs and giants''. However, he does not fit $\alpha$ as a free parameter and measure the uncertainty, so it is unclear what `well represented' here means. Furthermore for the later-type stars the profile fits are not compelling. It is noteworthy that the measured scale heights are much smaller than the canonical value for the thin disk of 300\,pc, even for the later-type stars for which one might expect agreement.  

The most detailed information on the vertical density distribution is provided by the recent study by \citet{Xiang2018} using LAMOST spectroscopic observations. They are able to divide stars into several age bins. The measured values of $\alpha=2/n_1$ (their Table 3b) show significant differences between age bins, both up and down, but averaging over several bins one can see that the younger populations, ages $<8$\,Gyr, have smaller scale heights and average $\alpha\sim1$, while the older populations, ages $>8$\,Gyr, have larger scale heights and $\alpha\sim0$. Combining all ages together the best fit value is also $\alpha\sim0$. The selection function for this survey is exceedingly complex, and the estimation of $\alpha$ was not a primary aim of the project. The significant variations in $\alpha$ between different age bins may indicate that the uncertainties have been underestimated, which is why we have not quoted uncertainties here. Nevertheless the overall trend of $\alpha$ decreasing with age seems clear and this is the first time that this has been shown.

In their own study, \citet{Dobbie} used the large samples of K and M stars from SDSS collated by \citet{1Ferguson} to study the problem. They found that there is moderate evidence (specifically meaning $2<\ln B<5$, where $B$ is  the Bayes factor) for the exponential and $\sech$ models over the $\sech^2$ model, but concluded that a sample that reaches closer to the Galactic plane is needed. This is in fact the problem with the majority of samples of the vertical structure of the Galactic disk, that they sample a conical volume, with the Sun at the  apex, so the space density at small heights $|z|<300$\,pc is not sufficiently well sampled, if at all. This may be compounded by the problem that the images of nearby stars are saturated. 

The results on $\alpha$ of \citet{Xiang2018} and \citet{Dobbie} are not in agreement with the finding of \citet{Bovy2017} that $\alpha\sim2$. As with the study of \citet{Xiang2018}, the selection function for the sample of \citet{Bovy2017} is rather complex. This makes it very difficult to investigate the origin of the disagreement. These results motivate a new measurement of $\alpha$ using a survey with good sampling of the local volume, distances $<500$\,pc, with a simple selection function, and ideally selected at near-infrared wavelengths to minimise extinction. In this paper we analyse such a sample: we combine 32\,942 M7 to M9.5 dwarfs from \citet{Ahmed2019} with 1\,016 L0 to L2.5 dwarfs from \citet{Skrzypek2016}, all selected from UKIDSS, to measure an accurate value of $\alpha$. This in turn provides a measurement of the space density of each spectral type in the plane of the disk, which can be transformed to the luminosity function at the bottom of the main sequence.

\begin{figure*}
\centering
\includegraphics[width = 15.cm, trim = 0cm 10cm 0cm 9cm]{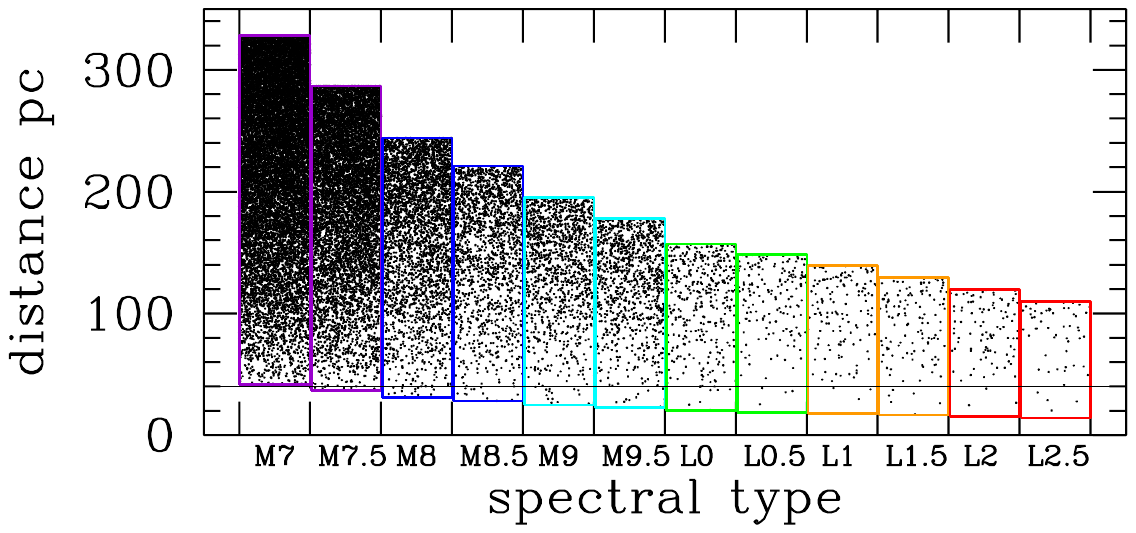}
\caption{Distribution of distances for the sample of 33\,985 M7 to L2.5 dwarfs, where for this plot sources are treated as single (unresolved binaries are ignored). Objects have been plotted at a random horizontal coordinate within each spectral-type box. For each spectral type the upper and lower distance limits correspond to the sample magnitude limits $13.0<J<17.5$. The sample is complete over all spectral types between distances $d_\mathrm{min}$(M7) and $d_\mathrm{max}$(L2.5), which is the range $41.3<d<109.5$\,pc. The thin horizontal line is drawn at a distance of 40\,pc, to make clear the relative contributions of the different spectral sub-types to measuring the space density at very small distances.}
\label{fig:sample}
\end{figure*}

The layout of the remainder of the paper is as follows. In Sect. \ref{sec:samples} we describe the samples used in the analysis. In Sect. \ref{sec:analysis} we analyse the vertical density distribution using firstly a binned estimate, and then a maximum-likelihood fit using every star individually. We provide a discussion of these results on the vertical density distribution in Sect. \ref{sec:discussion}. In Sect. \ref{sec:lf} with use the results to determine the stellar luminosity function over the spectral range M7 to L2.5.  We provide a summary of the main points in Sect. \ref{sec:summary}. All magnitudes in this paper are on the Vega system. The $J$ band refers to the MKO $J$ passband \citep{Tokunaga2002}, unless specifically stated otherwise.

\section{Samples}\label{sec:samples}

By combining the samples of \citet{Ahmed2019} and \citet{Skrzypek2016} we create a large homogeneous sample of 33\,958 M7 to L2.5 dwarfs $13.0<J<17.5$, at distances $<350$\,pc (except for unresolved binaries as explained below). The hydrogen burning limit is believed to be reached after spectral type L2.5 \citep{Dieterich2014}, meaning that L3 dwarfs and later are brown dwarfs, while L2.5 dwarfs and earlier are predominantly main sequence stars, but can also include young brown dwarfs. In the field the proportion of young brown dwarfs is small, and therefore the sample of M7 to L2.5 dwarfs is representative of the bottom of the main sequence. The numbers of dwarfs of different spectral types are listed in Table \ref{tab:counts}.
As detailed in the above catalogue papers these samples are well suited to measuring the density profile and the luminosity function. The samples are highly complete, and the spectral classifications are unbiased except for rare peculiar blue or red sources, comprising an estimated $\sim1\%$ of the sample.

\begin{figure}
\centering
\includegraphics[width = 9.cm, trim = 1cm 7cm 0cm 1cm]{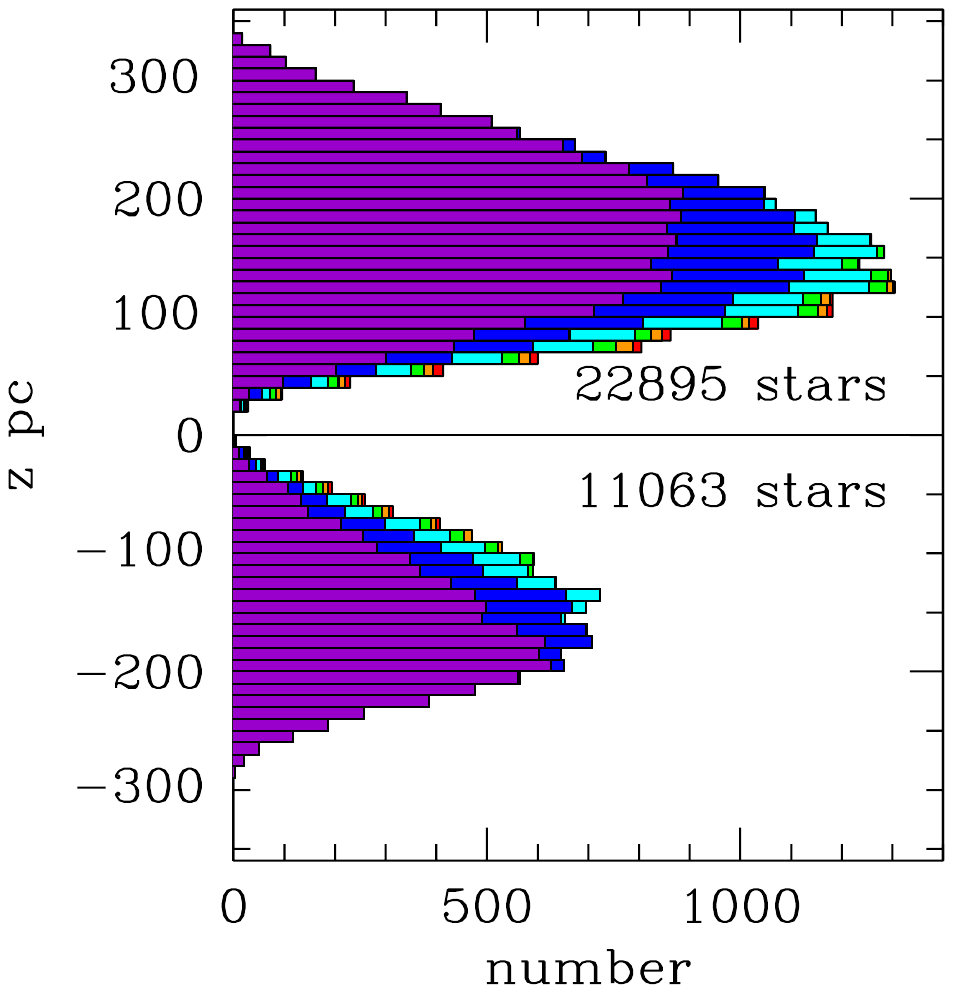}
\caption{Distribution of height $z$ above the Galactic plane for the sample, coloured by spectral type: M7+M7.5 (purple), M8+M8.5 (blue), M9+M9.5 (cyan), L0+L0.5 (green), L1+L1.5 (orange), L2+L2.5 (red). For this plot the Sun is assumed to lie at a height of 10\,pc above the plane.}
\label{fig:zheight}
\end{figure}

The two samples cover the same area of sky and were selected by essentially identical methods using the {\em phototype} method of \citet{Skrzypek2015}. 
The method classifies objects using multicolour photometry. For the L dwarfs the bands $izYJHKW1W2$ were used, while for the M dwarfs the W1 and W2 bands were omitted. They add no significant useful information for these types. The accuracy of the spectral types is competitive with spectroscopy. For the M dwarfs the classification is accurate to better than 0.5 sub-types {\em rms}, and is tied to the optical spectroscopy of the BOSS sample of \citet{Schmidt2015}. For the L dwarfs the classification is accurate to one sub-type {\em rms}, and is anchored to the optical system of \citet{Kirkpatrick1999}. The sample is presented in Fig. \ref{fig:sample}, where each star has been plotted at the distance computed assuming that the object is single (ignoring unresolved binaries). Distances are computed based on the absolute magnitudes listed for each spectral type in Table \ref{tab:counts}. The selected absolute magnitudes are discussed extensively below, in this section. The sample is plotted in a different way in Fig. \ref{fig:zheight}, as a histogram showing heights from the Galactic plane, assuming the Sun lies at a height of 10\,pc above the plane. This plot illustrates the fact that the sample includes a large number of objects at heights $|z|<100$\,pc, and therefore is well suited to investigating the softening close to the Galactic plane. The survey covers an effective area of 3\,031\,deg$^2$ ($7.3\%$ of the sky), and the solid angle as a function of Galactic latitude $\Omega(b)$ is provided in Table 1 of \citet{Ahmed2019}, in wedges of angular extent $1^{\circ}$. The sample defined in this way has excluded from the slightly larger total sample two small areas of larger reddening, totalling 39\,deg$^2$. 

\begin{table}
\centering
\caption{Number of sources and absolute magnitude \\ by spectral type}
\begin{tabular}{lrr|lrr}
\hline\hline
{SpT} & {Count} & $M_J$ & {SpT} & {Count} & $M_J$ \\ \hline
M7    & 15\,772 & 9.92 & L0 & 408 & 11.52 \\
M7.5 & 9\,242 & 10.21 & L0.5 &  175 & 11.64 \\
M8    & 3\,637 & 10.56 & L1 &  143 & 11.78 \\
M8.5 & 1\,851 & 10.78 & L1.5 &  130 & 11.94 \\
M9    & 1\,354 & 11.05 & L2 &  103 & 12.11 \\
M9.5 & 1\,086 & 11.25 & L2.5 &  57 & 12.30 \\ \hline
total & 32\,942 &   & total & 1\,016 & \\ \hline
\end{tabular}
\label{tab:counts}
\end{table}

The photometry for this sample is very precise and extinction is very low. In the $J$ band the median photometric uncertainty is 0.016, and the $90\%$ quantile is 0.028. We quantify the extinction using the results from \citet{Green2018}, for a distance of 400\,pc. This is an overestimate of the effects of dust on the sample, but the method of  \citet{Green2018} becomes too inaccurate at smaller distances to be useful. On this basis in the $J$ band the median extinction is $<0.02$\,mag. and for $90\%$ of sources the extinction is $<0.06$\,mag. For only $0.3\%$ of the sources is the extinction at 400\,pc greater than 0.13\,mag. Therefore we make no corrections for extinction.

\citet{Laithwaite2020} have made a detailed study of the unresolved binaries in the \citet{Ahmed2019} sample of M7 to M9.5 dwarfs. They find that unresolved binaries comprise $16.2\%$ of all the systems, and that the binaries are almost exclusively equal mass systems. This means that we can assume that per unit volume $16.2\%$ of the sources are twice as bright as single stars. We will assume that the same properties apply to the L0 to L2.5 dwarfs. In this respect we note that \citet{Reid2008} find a similar fraction of unresolved binaries in their sample of L dwarfs.

\citet{Laithwaite2020}  also redetermined values of the absolute magnitude $M_J$ for spectral types M7 to M9.5, based on {\em Gaia} DR2 parallaxes for 2706 systems, with typical parallax/error of 30.\footnote{We have checked these results using the more recent {\em Gaia} EDR3 release, finding essentially identical values.} The results therefore are very accurate, but the values are on average some 0.5\,mag. brighter than those of \citet{Dupuy2012} for these spectral types (although the two relations converge at L0). The new values are listed in Table \ref{tab:counts}. The distances quoted in \citet{Ahmed2019} used the absolute magnitudes of \citet{Dupuy2012} and are therefore wrong.

The large difference in $M_J$ for spectral types M7 to M9.5 compared to  \citet{Dupuy2012} requires further comment. The matter was considered in detail by \citet{Laithwaite2020}, in their Section 6.1.
The reason for the discrepancy is not due to incorrect parallaxes in \citet{Dupuy2012} but is apparently due to differences in spectral classifications between the two samples. The sample of \citet{Ahmed2019} is homogeneous and the classifications are accurately calibrated to the classifications of the BOSS spectroscopic sample of \citet{Schmidt2015}, which itself is homogeneous and was subject to careful checks for systematics. \citet{Laithwaite2020} found 10 stars in common between \citet{Schmidt2015} and \citet{Dupuy2012}, and the classifications of \citet{Schmidt2015} are on average 0.7 spectral types later. 

Some additional evidence that the \citet{Schmidt2015} M7 to M9 classifications are later than older classifications comes from the recent CARMENES paper by \citet{Cifuentes}. Although there are no CARMENES M7 to M9 stars in common with \citet{Schmidt2015}, it is evident that over this spectral range the $G-J_{2MASS}$ colours in \citet{Cifuentes} are substantially redder than the $G-J_{MKO}$ colours in \citet{Laithwaite2020}. Over this spectral range a star has $J_{\mathrm 2MASS}-J_{\mathrm MKO}\sim0.1$\,mag. \citep{Stephens2004}. Using this, their average colours for M7 and M8 are $G-J_{MKO}=4.15, 4.42$ respectively. \citet{Laithwaite2020} measure $G-J_{MKO}=4.15, 4.47$ for spectral types M8 and M9, respectively, indicating an offset by approximately one spectral type. The absolute magnitude scale of \citet{Cifuentes} is also quite close to that of \citet{Dupuy2012}.

To summarise, there is now good evidence that the differences in absolute magnitude between the relation of  \citet{Laithwaite2020} and the other two samples are due to differences in spectral classifications between the BOSS sample of \citet{Schmidt2015} and older samples. The advantage of using the M7 to M9.5 sample of \citet{Ahmed2019} is that it is large and homogeneous, and the relation between $M_J$ and spectral type has been accurately calibrated with {\em Gaia} with a large sample of 2706 systems. It would of course be incorrect to apply the relation between $M_J$ and spectral type derived from a different sample to the sample of \citet{Ahmed2019}.
The sample of \citet{Schmidt2015}, to which the spectral types of \citet{Ahmed2019} are calibrated, has become the {\em de facto} standard in this field, so this apparent offset in spectral classifications compared to older samples needs to be borne in mind when comparing science results derived from different samples.

For the L0 to L2.5 stars the absolute magnitudes in Table \ref{tab:counts} were calculated using the polynomial relation between $M_J$ and spectral type of \citet{Dupuy2012}, derived from ground-based parallaxes. We checked these values by first matching to {\em GAIA} DR2 all the L0 to L3  dwarfs in the sample of \citet{Skrzypek2016}, then limiting to sources with parallax/error$>10$. We then fit a linear relation between absolute magnitude and spectral type, allowing for binaries by fitting a double Gaussian profile to the distribution of absolute magnitudes at fixed spectral type. The method is very similar to that employed by \citet{Laithwaite2020} for the M7 to M9.5 stars, except they used $G-J$ colour rather than spectral type. The result for single stars is the linear relation $M_J=0.359\,\mathrm{SpT} + 7.882$ where SpT denotes spectral type and L0, L3 are 10, 13\footnote{Beware that others, e.g. \citet{Bardalez2019}, use 20, 23 for L0, L3.}. Comparing against the values in Table \ref{tab:counts} we find agreement at the level $|\Delta M_J|<0.1$ for all sub-types L0 to L2.5, confirming that the absolute magnitudes of \citet{Dupuy2012} are reliable over this spectral range.

A final point to consider is the possibility of bias in the distances, due to the application of spectroscopic parallaxes i.e. a single absolute magnitude for all the stars of a particular spectral type. This could lead to biases in the derived structural parameters i.e. the scale height $z_e$ and the value of $\alpha$.
We present an analysis of this point in the Appendix, finding that any systematic errors introduced are significantly smaller than the random errors on parameters.

\section{Fitting the vertical density distribution}\label{sec:analysis}

It is usual to measure the density distribution by first binning the data and then fitting to the counts. This is useful because the binned counts give a visual impression of the shape of the variation in density. There is an important drawback to this approach however, in that it is unclear how to deal with the unresolved binaries: in Fig. \ref{fig:sample} the binaries (which cannot be identified individually) should be plotted at a distance a factor $\sqrt{2}$ larger. Here we first present a binned analysis of the M7 and M7.5 stars only, which includes a large fraction of all the stars, and we then provide results of an optimal method that fits to all the data points simultaneously, without binning, and correctly accounts for binaries.

\subsection{Binned analysis}\label{sec:binned}

\begin{figure}
\centering
\includegraphics[width = 11.cm, trim = 1cm 7cm 0cm 3cm]{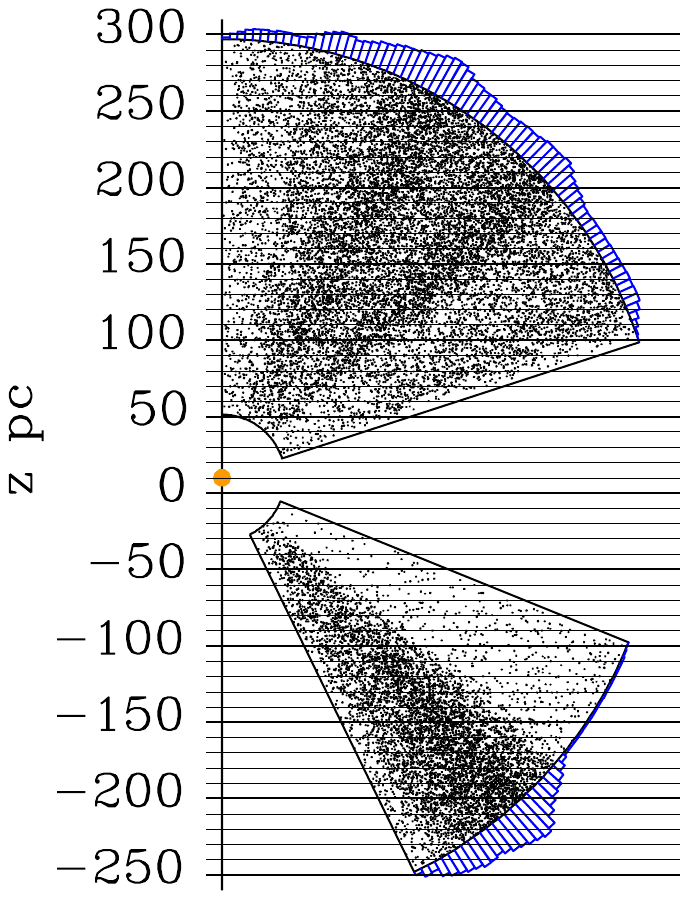}
\caption{Polar plot, with coordinates distance $d$ and Galactic latitude $b$, for the volume-complete sample of 20\,849 M7 and M7.5 stars with distances between $d_\mathrm{min}$(M7)$=41.3$\,pc and $d_\mathrm{max}$(M7.5)$=287.1$\,pc. The blue histogram shows the areal coverage, with the radial length of each bin proportional to the solid angle at that $b$. The orange dot indicates the observer, located 10\,pc above the Galactic plane. The horizontal slices are those used in the binned estimates of the space density.}
\label{fig:bin1}
\end{figure}

In this section we simply ignore the fact that a fraction of the sources are unresolved binaries, and treat all the sources as single. The results are illustrative and used as a guide to the more complete analysis presented in the next subsection. Referring to Fig. \ref{fig:sample} the lower and upper distance limits for each spectral type correspond to the magnitude limits of the survey $J=13.0$ and 17.5, given the absolute magnitude for any particular spectral type. Therefore we can form a volume-complete sample of M7 and M7.5 stars by using distance limits $d_{min}$(M7), and $d_{max}$(M7.5), which are 41.3\,pc and 287.1\,pc respectively. The sample comprises 20\,849 stars and is plotted in Fig. \ref{fig:bin1} using polar coordinates. The blue histogram plots the solid angle of the survey at each value of $b$, in $1^\circ$ wedges. Therefore to compute the space density we sum the number of sources in each slice, of height 10\,pc, and sum the volume contributed by each wedge along the slice, accounting for the solid angle as it varies with $b$. For this calculation we assume the Sun lies at a height $z_\odot=10$\,pc above the Galactic plane. 

\begin{figure}
\centering
\includegraphics[width = 9.cm, trim = 1cm 8cm 0cm 4cm]{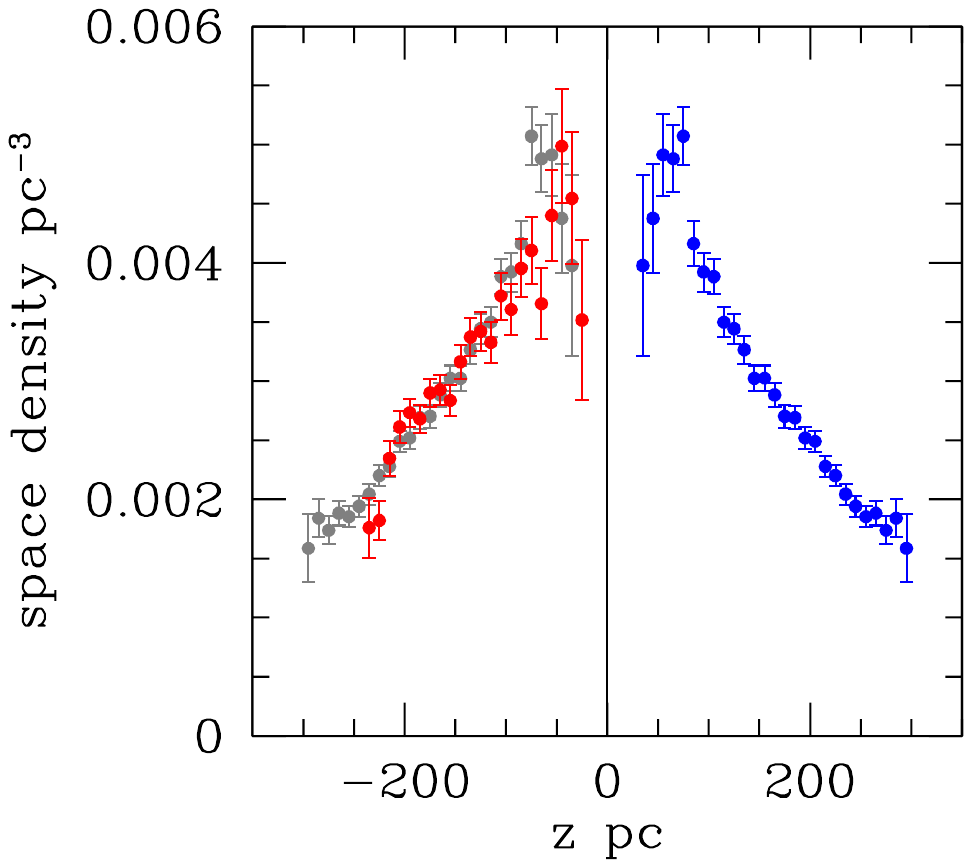}
\caption{Binned estimate of the variation of space density with height $z$ above the Galactic plane, for the volume-complete sample of 20\,849 M7 and M7.5 stars with distances $41.3<d<287.1$. Red points - below the plane; blue points - above the plane; grey points - the blue points reflected through the origin.}
\label{fig:bin2}
\end{figure}

\begin{figure}
\centering
\includegraphics[width = 9.cm, trim = 1cm 8cm 0cm 4cm]{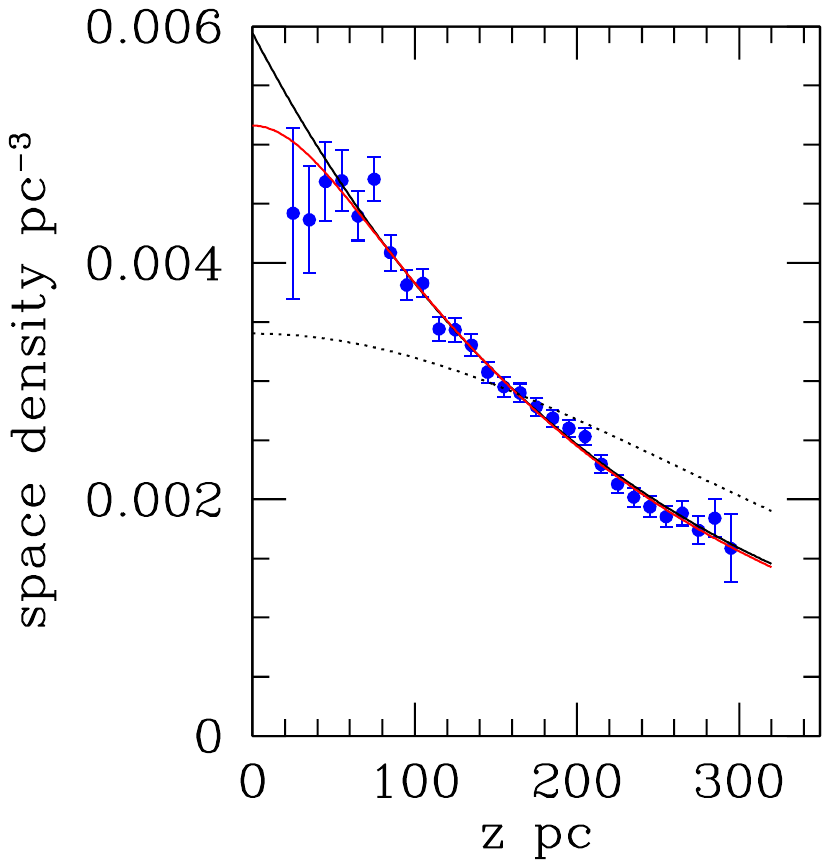}
\caption{Binned estimate of the variation of space density with absolute height $|z|$ above the Galactic plane, for the volume-complete sample of 20\,849 M7 and M7.5  stars with distances $41.3<d<287.1$\,pc. Red curve: best fit $\sech^\alpha$ with $\alpha=0.23$. Black curve: best fit exponential. Dotted curve: best fit sech$^2$ with scale height $z_e$ fixed to 200\,pc.}
\label{fig:bin3}
\end{figure}

The results of this calculation are plotted in Fig. \ref{fig:bin2}. The blue points are the binned estimates of space density at heights above the plane, and the red points are the same for below the plane. The uncertainty on each point is plotted as a fractional uncertainty of $1/\sqrt{N}$, where $N$ is the number of points in the slice, and we have only used bins with $>20$ points. The grey points are the blue points reflected about the Galactic plane, and allow a comparison of the variation in space density above and below the plane. There are no strong differences between the two curves, indicating consistency. It is well known that at larger distances from the plane differences are seen when comparing measurements above and below the plane \citep[e.g.][]{Widrow2012,1Ferguson}. It is possible that density fluctuations exist at a similar level in our data but they would be relatively less important close to the Galactic plane where the space densities are higher. Given the good agreement between the red and the grey points we are justified in averaging the results for above and below the plane. The averaged results are presented in Fig. \ref{fig:bin3}. 

In fitting a model to the data, we assume Gaussian errors for each bin i.e. we approximate the Poisson distribution as Gaussian. Then if the model predicts $m_i$ points in bin $i$, and the observed number is $n_i$, the logarithm of the likelihood is given by:
\begin{equation}
\ln\mathcal{L} = -\sum\frac{(n_i - m_i)^2}{2m_i} - \frac{1}{2}\sum\ln m_i \:,
\label{eqn:loglike}
\end{equation}
which is Eqn 8 in \citet{Dobbie}.
Because the likelihood is strongly peaked in the space of the parameters, the posterior is not sensitive to the form of the priors. We adopt broad uniform priors, which for the shape parameter covers the range $0<\alpha<3$. The function $\sech^{\alpha}(z/\alpha z_e)$ can be awkward\footnote{this comes from overflow in evaluating $e^{(z/\alpha z_e)}$} to evaluate for small values of $\alpha$, so we employ the identity 
\begin{equation}
\sech^{\alpha}(x/\alpha)=2^{\alpha}e^{-x}(1+e^{-2x/\alpha})^{-\alpha} \:,
\end{equation}
and ensure that when $\alpha=0$ the function is set equal to $e^{-x}$.

Fitting Eqn \ref{eq:war}, we measure $\alpha=0.23\pm0.13$, and $z_e=222^{+8}_{-10}$\,pc. This quantifies that the density profile is consistent with exponential all the way to the Galactic plane and that any flattening is modest.\footnote{We also confirmed that the measured value of $\alpha$ is not very sensitive to the chosen value of $z_\odot=10$\,pc. For $z_\odot=20$\,pc we find $\alpha=0.31\pm0.12$.} The best-fit profile is plotted in Fig. \ref{fig:bin3}, where it is compared to the best-fit exponential, which has a scale height $z_e=227^{+7}_{-6}$\,pc. This indicates that any softening is confined to within 50\,pc of the plane. The data are inconsistent with not only the $\sech^2$ profile, but also the $\sech$ profile. Interestingly fitting a $\sech^2$ profile, which is a bad fit, yields a scale height of 126\,pc which makes no sense when compared to the fiducial scale height of the thin disk of 300\,pc. This calculation shows that force-fitting the wrong profile, one that flattens too much in the centre, leads to an underestimate of the correct scale height. The problem of the $\sech^2$ profile may be illustrated in another way. Because we have not accounted for binaries the binned analysis underestimates the true scale height somewhat. If we suppose that the scale height, uncorrected for binaries, could be as small as $z_e=200$\,pc, and fit a $\sech^2$ profile with this scale height the result is the very flattened and clearly incorrect profile plotted as the dotted line in Fig. \ref{fig:bin3}.

This binned analysis indicates that the profile is peaky near the centre, but the values of the fitted parameters are not correct because we have ignored the presence of unresolved binaries. It is well know that the presence of unresolved binaries in a sample causes the scale height to be underestimated, and conceivably it could affect the estimate of $\alpha$ as well. The effect of unresolved binaries is usually quantified by modelling \citep[e.g.][]{Covey08,1Juric,1Bochanski}. For example in the current case a synthetic catalogue would be created, one that includes unresolved binaries and that matches the selection criteria of the actual catalogue. The catalogue would be created with a scale height that is a guess for the true scale height and would be analysed in the same way as the actual sample. The catalogue scale height would then be adjusted until the measured (biased) scale height matches the measured value in the actual catalogue. In this way the true scale height is recovered.

In the next subsection we describe a likelihood  analysis that instead accounts for unresolved binaries in a direct way, and makes additional improvements over the binned analysis presented in this section.

\subsection{Full likelihood analysis}\label{sec:full}

\subsubsection{Method}\label{sec:method}

We now implement four improvements compared to the above binned analysis. These are:
\begin{enumerate}
\item Using stars individually without binning.
\item Accounting correctly for the presence of unresolved binaries.
\item Accounting for the intrinsic spread in absolute magnitude $M_J$ of each spectral type, which in a magnitude-limited sample means that intrinsically brighter sources are over-represented \---\ the Malmquist bias.
\item Including all stars in the sample from M7 to L2.5 over the full distance range of each type (displayed in Fig. \ref{fig:sample}).
\end{enumerate}

We wish to calculate the likelihood $\mathcal{L}$ of observing the sample in question. In deriving this we will assume a Poisson point process, and follow a similar procedure to that presented by \citet{Marshall1983}. Consider firstly a single spectral sub-type $t$. The data comprise the list of sources of observed Galactic latitude $b$ and apparent magnitude $J$. Ignoring binaries, the expected number of sources $\mu$ in an infinitesimal element $dbdJ$ is:
\begin{equation}
\mu=\rho(z(d(J),b,z_\odot))d^2(J)\Omega(b)db\frac{dd}{dJ}dJ\:.
\label{eqn:expect}
\end{equation}
Here, and throughout this section, $d(J)$ refers to the distance of a single star computed from $J$ using the absolute magnitude for the particular sub-type, listed in Table \ref{tab:counts}. The height above the plane is calculated as $z(d(J),b,z_\odot)=d(J)\sin(b)+z_\odot$. The angle $b$ is in degrees, since $\Omega(b)$ is the solid angle defined in wedges of angular size $1^\circ$ (see Sect. \ref{sec:samples}). The term $\rho(z(d(J),b,z_\odot))$ should be understood to include the dependence on the model parameters $\rho_0(t),$ $z_e$ and $\alpha$. 

The probability of finding one source in the element is $\mu e^{-\mu}$, and of finding none is $e^{-\mu}$. Therefore the likelihood is the product of the probabilities of observing the $N$ sources with their particular $b,J$, and of observing no sources in all the other elements.  Consequently the likelihood for this type $t$ may be written:
\begin{equation}
\mathcal{L}_t = \prod_i \mu_i e^{-\mu_i}\prod_j e^{-\mu_j} \:,
 \label{eqn:likelihood}
\end{equation}
where the first product is over elements containing sources, and the second product is over all the other elements within the volume surveyed.
Taking the logarithm and dropping terms that are independent of the model parameters we obtain:
\begin{equation}
\ln\mathcal{L}_t = \sum_i \ln(\rho(z(d(J_i),b_i,z_\odot))-\iiint_V\rho dV \:,
 \label{eqn:likelihood2}
\end{equation}
and the volume integral is the expected number of sources in the survey, given the density function $\rho$, which depends on $\rho_0(t), z_e, \alpha$
and $z_\odot$. The likelihood as defined above uses all the stars of a particular sub-type individually, and deals with the first item above. 

We now consider the treatment of binaries. It is relatively easy to treat the binaries correctly through the likelihood because we can assume that each binary is exactly twice as bright as a single, based on the detailed study of binaries in this sample by \citet{Laithwaite2020}. This means that the survey is in fact two surveys, one for singles, and a second for binaries, where the distance limits for the binary survey are $\sqrt{2}$ larger than for the singles survey. 
For a total number of systems comprising a fraction $f_b$ of unresolved equal-mass binaries, and so a fraction $1-f_b$ singles, if the space density of stars at any point is $\rho$, the space density of single stars is $\rho(1-f_b)/(1+f_b)=\rho r_s$, and the space density of binary systems $\rho f_b/(1+f_b)=\rho r_b$. The binary systems are unresolved sources that are twice as bright as single stars, and we assume $f_b=0.162$ \citep{Laithwaite2020}. 

Returning now to equation \ref{eqn:expect}, in computing the expected number of sources in an element $dbdJ$ we must include the expected number of binary systems, and the equation becomes:
\begin{equation}
\begin{split}
\mu=[r_s\rho(z(d(J),b,z_\odot))+2\sqrt{2}r_b\rho(z(\sqrt{2}d(J),b,z_\odot))]\\ \times d^2(J)\Omega(b)db\frac{dd}{dJ}dJ\:,
\end{split}
\label{eqn:expect2}
\end{equation}
where the term $2\sqrt{2}$ in front of $r_b$ derives from the larger volume element at the larger distance (from the terms $d$ and $dd/dJ$).
Propagating through to the logarithm of the likelihood, we obtain the final expression
\begin{equation}
\begin{split}
\ln\mathcal{L}_t = \hspace{7.5cm}
\\ \sum_i \ln[r_s\rho(z(d(J_i),b_i,z_\odot))+2\sqrt{2}r_b\rho(z(\sqrt{2}d(J_i),b_i,z_\odot))] \\ -\iiint_Vr_s\rho dV_s-\iiint_Vr_b\rho dV_b \:,
\end{split}
 \label{eqn:likelihood3}
\end{equation}
where the first triple integral is over the volume occupied by singles, given the sample magnitude limits, and the second volume integral is the same for the binary systems, for which all distances are $\sqrt{2}$ larger. In this way binaries are correctly included in the calculation of the likelihood.

We now consider how to treat Malmquist bias. Our sample is magnitude limited $13.0<J<17.5$. But (single or binary) stars of a particular spectral type have a spread in absolute magnitude (due to e.g. variations in age and/or metallicity). Therefore the more luminous sources are detected to larger distances, and are overrepresented in the sample, and {\em vice versa} for less luminous sources. If stars of a particular sub-type are treated as having a unique absolute magnitude the measured parameters of the density distribution will be biased, and this is what we mean when using the term Malmquist bias in this paper. \citet{Laithwaite2020} found a Gaussian distribution of absolute magnitude of dispersion $\sigma_M=0.21$\,mag. at fixed $G-J$ colour. Over the colour spread of half a spectral sub-type the dispersion increases to $\sigma_M=0.24$\,mag. which is the value we adopt. The additional dispersion comes from the relation between absolute magnitude and colour.

Eqn \ref{eqn:likelihood3} shows us how to deal with the spread in the absolute magnitudes $M_J$ of each spectral type. In Eqn \ref{eqn:likelihood3} we are dealing with two populations, singles and binaries, where the binaries are twice as bright and occupy a different volume to the singles, where all distances are $\sqrt{2}$ larger. In the same way each of these two populations comprises a set of sub-populations of different absolute magnitude, the more luminous sources occupying a volume where all distances are multiplied by a factor $f>1$, compared to the average, and the less luminous sources occupying a volume where all distances are multiplied by their own factor $f<1$. To implement this we divide each population (singles or binaries) into a small number of sub-populations, i.e. we model the Gaussian distribution of $M_J$ as a coarse histogram. Each sub-population of absolute magnitude $M_J-\Delta M_J$ has a distance correction $f=10^{0.2\Delta M_J}$ and a volume correction $f^3$, analogous to the $\sqrt{2}$ and $2\sqrt{2}$ terms in the second term in Eqn \ref{eqn:likelihood3}. For a set of subpopulations defined by weights $w_j$ $(\sum_j w_j=1)$ and distance corrections $f_j$, then, for example, the first term in Eqn \ref{eqn:likelihood3}, $r_s\rho(z(d(J_i),b_i,z_\odot))$, is replaced by $r_s\sum_j w_jf_j^3\rho(z(f_jd(J_i),b_i,z_\odot))$. There is a similar sum for the second term, and then a set of volume integrals for all the sub-populations, single as well as binary, over the relevant volume occupied by each sub-population.

In principle photometric errors can have an effect that is similar to the effect of the spread in absolute magnitudes, but this can be safely ignored for this dataset as the photometric errors in the $J$ band (Sect. \ref{sec:samples}) are considerably smaller than the dispersion in absolute magnitude.

The final improvement we make ensures that all the stars in the full sample are used, over the full distance range of each spectral sub-type (see Fig. \ref{fig:sample}), rather than limiting to the distance range in common, as we did in the binned analysis for the M7 and M7.5s. This can be achieved straightforwardly by assuming that the density function has the same form for each spectral sub-type, meaning that the parameters $z_e,\alpha,z_\odot$ are in common, but the normalisations are different, i.e. the central space density of each spectral type is a free parameter $\rho_0(t)$. Then the likelihood is the sum of the individual likelhoods for each sub-type, $\ln\mathcal{L}=\sum_t \ln\mathcal{L}_t$, where the individual likelihoods are computed over the full sample and full volume for that sub-type. This means that the total number of free parameters is 15: the 12 $\rho_0(t)$, and $z_e,\alpha,z_\odot$. To be completely clear: $\rho_0(t)$ is the summed number of stars (not systems) per unit volume for a particular sub-type.

We adopt broad uniform priors on the parameters. Again, because the likelihoods are sharply peaked, the results are insensitive to the priors. 

\begin{table}
\centering
\caption{Best fit values and their uncertainties for the full likelihood analysis}
\begin{tabular}{lcc}
\hline\hline
parameter & \multicolumn{1}{c}{$\sech^\alpha$} & \multicolumn{1}{c}{exponential}  \\ \hline
$\rho_0\,$M7\:\:\:\:\:\:\:\:\:pc$^{-3}$   & $ 2.37^{+0.14}_{-0.09}\times10^{-3} $ & $ 2.79^{+0.06}_{-0.05}\times10^{-3} $ \\
$\rho_0\,$M7.5 & $ 1.88^{+0.12}_{-0.07}\times10^{-3} $ & $ 2.22^{+0.04}_{-0.04}\times10^{-3} $ \\
$\rho_0\,$M8   & $ 1.09^{+0.06}_{-0.04}\times10^{-3} $ & $ 1.28^{+0.03}_{-0.03}\times10^{-3} $ \\
$\rho_0\,$M8.5 & $ 0.71^{+0.05}_{-0.03}\times10^{-3} $ & $ 0.84^{+0.02}_{-0.02}\times10^{-3} $ \\
$\rho_0\,$M9   & $ 0.71^{+0.04}_{-0.03}\times10^{-3} $ & $ 0.83^{+0.03}_{-0.02}\times10^{-3} $ \\
$\rho_0\,$M9.5 & $ 0.72^{+0.05}_{-0.03}\times10^{-3} $ & $ 0.85^{+0.03}_{-0.03}\times10^{-3} $ \\ \hline
$\sum\rho_0\,$M7-M9.5   & $ 7.48^{+0.44}_{-0.28}\times10^{-3} $ & $ 8.80^{+0.15}_{-0.15}\times10^{-3} $ \\ \hline
$\rho_0\,$L0   & $ 0.38^{+0.03}_{-0.02}\times10^{-3} $ & $ 0.44^{+0.02}_{-0.02}\times10^{-3} $ \\
$\rho_0\,$L0.5 & $ 0.19^{+0.02}_{-0.02}\times10^{-3} $ & $ 0.22^{+0.02}_{-0.02}\times10^{-3} $ \\
$\rho_0\,$L1   & $ 0.18^{+0.02}_{-0.02}\times10^{-3} $ & $ 0.21^{+0.02}_{-0.02}\times10^{-3} $ \\
$\rho_0\,$L1.5 & $ 0.20^{+0.02}_{-0.02}\times10^{-3} $ & $ 0.23^{+0.02}_{-0.02}\times10^{-3} $ \\
$\rho_0\,$L2   & $ 0.20^{+0.02}_{-0.02}\times10^{-3} $ & $ 0.23^{+0.02}_{-0.02}\times10^{-3} $ \\
$\rho_0\,$L2.5 & $ 0.14^{+0.02}_{-0.02}\times10^{-3} $ & $ 0.16^{+0.02}_{-0.02}\times10^{-3} $ \\ \hline
$\sum\rho_0\,$L0-L2.5   & $ 1.29^{+0.08}_{-0.06}\times10^{-3} $ & $ 1.50^{+0.05}_{-0.05}\times10^{-3} $ \\ \hline
$\sum\rho_0\,$M7-L2.5   & $ 8.77^{+0.51}_{-0.32}\times10^{-3} $ & $ 10.30^{+0.16}_{-0.16}\times10^{-3} $ \\ \hline
$z_e$\,pc       & $ 258.6^{+10.2}_{-12.3}$ & $269.3^{+6.6}_{-6.3}$ \\ 
$\alpha$   & $ 0.29^{+0.12}_{-0.13} $ & ... \\ 
$z_\odot$\,pc & $ 10.9^{+1.7}_{-1.6} $ & $ 10.0^{+1.5}_{-1.4} $ \\ \hline
\end{tabular}
\label{tab:results}
\end{table}

\subsubsection{Results}

We have fit the function $\sech^\alpha$, as well as the simpler exponential function. We used the MCMC package {\tt emcee} \citep{Foreman2013} to maximise the likelihood, and to measure the uncertainties. The results for both functions are summarised in Table \ref{tab:results}, in each case accounting for binaries and  Malmquist bias in the analysis. The uncertainties quoted correspond to the $16, 50, 84\%$ quantiles in the marginilised distributions, and for all the $\rho_0(t)$ the fractional uncertainties are larger than $1/\sqrt{N}$. The uncertainties on the $\rho_0(t)$ parameters are considerably larger for the $\sech^\alpha$ function compared to the exponential function because they include the uncertainty in the flattening towards the plane. We use the $\sech^\alpha$ results when comparing against local measurements of space densities, and in calculating the luminosity function. The exponential function fit is included because of its simplicity, and it can be used for comparison against other surveys except close to the plane $|z|<50$\,pc.

The uncertainties on the $\rho_0(t)$ parameters are highly correlated. Therefore when we perform arithmetic on the space densities (e.g. the summed space density of spectral types M7-M9.5 listed in Table \ref{tab:results}, and later the calculation of the luminosity function), to measure the uncertainties we first perform the arithmetic on the MCMC chains and then measure the dispersion in the resulting values. 

\begin{figure}
\centering
\includegraphics[width = 8.5cm, trim = 0cm 0cm 0cm 0cm]{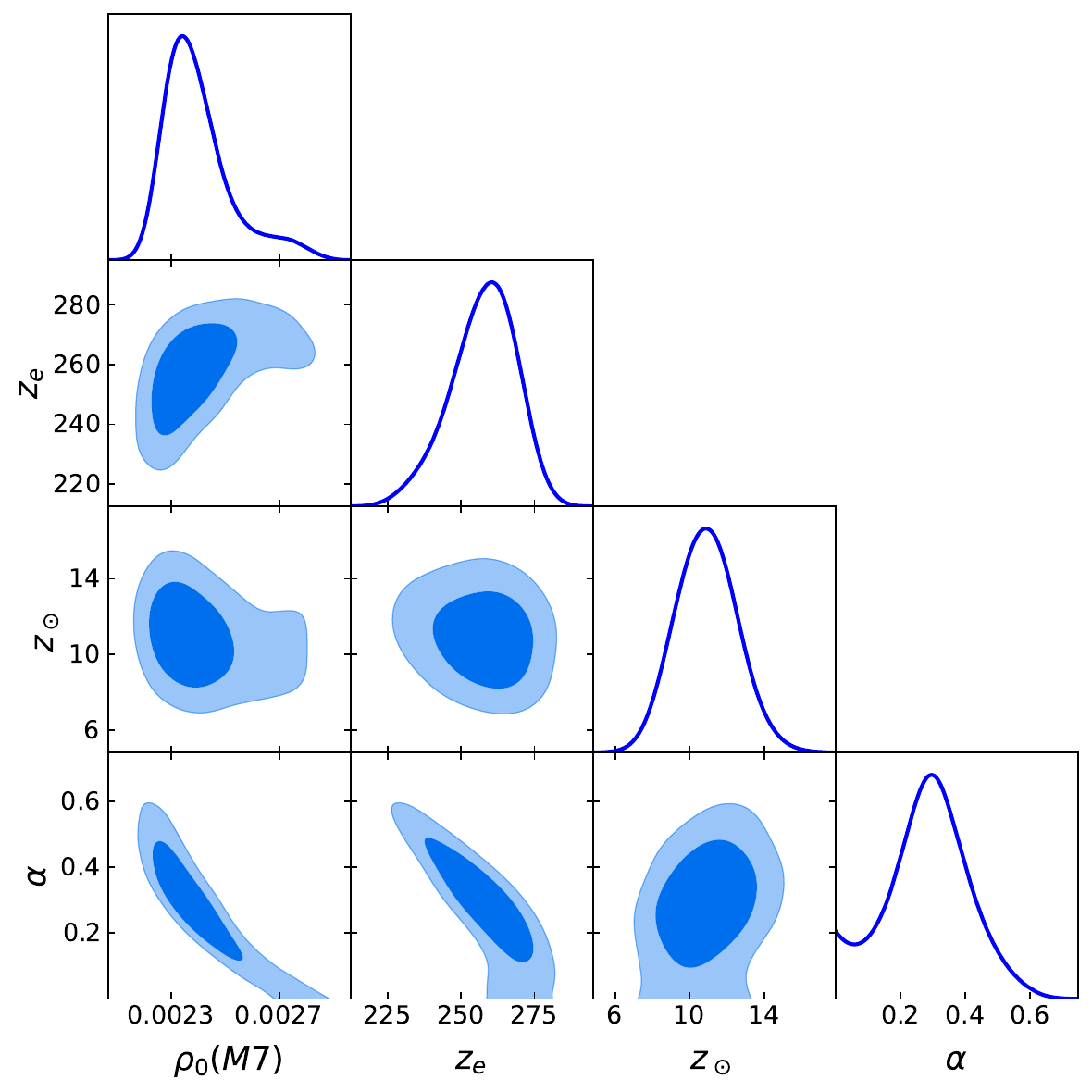}
\caption{Corner plot of the posterior probability density for the parameters: central density $\rho_0$ for the M7s, the scale height $z_e$, the offset of the Sun from the plane $z_\odot$, and the shape parameter $\alpha$. The contours contain $68\%$ and $95\%$ of the posterior probability.}
\label{fig:corner}
\end{figure}

A corner plot for the parameters $\rho_0$ for the M7s, $z_e$, $z_\odot$, and  $\alpha$ is presented in Fig. \ref{fig:corner}, produced using the {\tt GetDist} package \citep{Lewis2019}.  We have included only one $\rho_0(t)$ in this plot as the correlations for $\rho_0(t)$ of the other spectral sub-types have a similar form. 

\begin{figure}
\centering
\includegraphics[width = 9.cm, trim = 1cm 8cm 0cm 4cm]{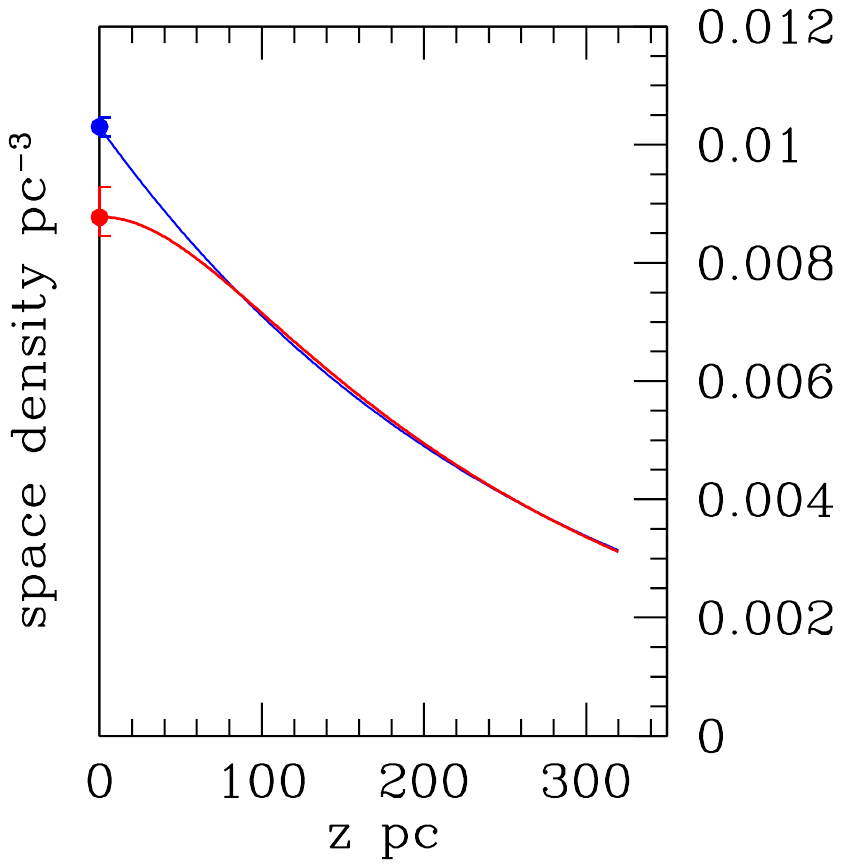}
\caption{Final best-fit density profiles from the likelihood analysis, summed over M7 to L2.5: $\sech^\alpha$ (red) and exponential (blue).}
\label{fig:bestfits}
\end{figure}

The most interesting result, and the principal result of this paper, is the distribution for $\alpha$. As listed in Table \ref{tab:results} we measure $\alpha=0.29^{+0.12}_{-0.13}$. Although the $\alpha$ distribution is strongly peaked near 0.3, there is a shoulder to the distribution that extends to $\alpha=0$. This shoulder reflects the fact that the data have almost no constraining power over the range $0<\alpha<0.1$ because within this range the density distribution varies only very close to the plane $|z|<20$\,pc, where there are very few sources, so the posterior is quite flat over this range. Quantifying the credible interval by the integrated probability within a range between equal probability densities, we find the $95\%$ and $99\%$ credible intervals are $0<\alpha<0.50$ and $0<\alpha<0.59$, meaning that the $\sech$ profile, $\alpha=1$, is firmly excluded. We wish to quantify at what level the credible interval includes the exponential model. This is ambiguous because the posterior density rises slightly as $\alpha$ approaches zero (Fig. \ref{fig:corner}). A useful measure is to state the credible interval at which the range becomes one sided, i.e. once the full range of $\alpha$ from the peak down to $\alpha=0$ is included, and this is the $95\%$ interval. This means there is moderate evidence against the exponential model continuing all the way to the Galactic plane, or equivalently moderate evidence for some degree of flattening close to the plane. We compare the exponential and $\sech^\alpha$ fits in Fig. \ref{fig:bestfits}, plotting the summed density for the full spectral range M7 to L2.5. The two curves are essentially identical except at heights $|z|<50$\,pc. Any softening of the exponential profile is rather slight. 

We can quantify the effects of the different improvements implemented in the full likelihood analysis. For the binned data we had the pair of results for the scale height $z_e$ and the shape parameter $\alpha$ of [222, 0.23]. Implementing successively a) treating all points individually rather than binned, and over the full distance ranges for each sub-type, b) accounting for binaries, c) accounting for Malmquist bias, these pairs become a) [227, 0.23], b) [252, 0.26], c) [259, 0.29]. We see that accounting correctly for binaries has a substantial effect. Without allowance for binaries the scale height is underestimated by $10\%$. The effect of Malmquist bias is considerably smaller. If not accounted for, the scale height is underestimated by $3\%$. 
We believe this is the first time that the corrections for binaries and for Malmquist bias have been made in this direct way, as opposed to using mock catalogues.

The sign and the size of these biases are in very good agreement with the results of \citet{1Juric} for the SDSS, computed using mock catalogues. For example they found (we find) that for a binary fraction $f_b=0.25\,(0.162)$ the scale height is underestimated by $15\%\,(10\%)$. For Malmquist bias they found (we find) that for $\sigma_M=0.30\,(0.24)$ the scale height is underestimated by $\sim5\%\,(3\%)$. 
 
Using each source individually rather than binning the data ensures that the data are used optimally. However, this has a large cost in terms of the computational time required for the fit, and the gain is actually probably rather modest. For much larger sample sizes than used here the computational cost could be prohibitive. In such cases a compromise is possible. The key would be to bin the data in $J$ and $b$, rather than in $z$, for each spectral sub-type. Then it would still be possible to implement a likelihood treatment for binaries and Malmquist bias analogous to the one employed here. 

\section{Discussion of the measured density profile}\label{sec:discussion}

The measured value of $\alpha=0.29^{+0.12}_{-0.13}$ for this population of stars corroborates the finding of \citet{Xiang2018} that $\alpha\sim 0$ for 
the thin disk, summing all ages together. The result is also in good agreement with the measurement by \citet{deGrijs} of the distribution of $\alpha$ in nearby edge-on spiral galaxies of $\alpha=0.5\pm0.2$. Since these latter measurements were made in the $K$ band the light would be dominated by cooler stars.\footnote{There is a caveat that the $K-$band light will be dominated by giants, whereas our study analyses the distribution of dwarfs. Based on the study of \citet{Rix93}, \citet{deGrijs} argue that the contribution of young red supergiants to the $K-$band light is small, and that the profiles of cool dwarfs and giants will be similar.} \citet{Bovy2017} found that the vertical profiles of A to K stars are `well represented by $\sech^2$ profiles', and he measured scale heights increasing from $\sim50$\,pc for (younger) A stars to $\sim150$\,pc for (older) G and K dwarfs. The latter value is much smaller than the canonical value for the thin disk of 300\,pc which is hard to understand. However in contrast to the A stars, the vertical profiles of G and K dwarfs are not well sampled by the GAIA DR1 data. It may be possible to reconcile all these results in the following way. The results of \citet{Xiang2018}  indicate that $\alpha$ is larger for young populations, so the profile for A stars might be satisfactorily fit by a $\sech^2$ profile. However the G and K samples will be dominated by older stars so for these populations one would expect a value of $\alpha$ similar to our measurement of 0.29. If this is true, fitting the $\sech^2$ profile to this steeper profile will result in a substantial underestimate of the scale height, as we found in Sect. \ref{sec:binned} (the same effect is also visible as the anti-correlation between $z_e$ and $\alpha$ in Fig. \ref{fig:corner}). This might help explain the small scale heights measured by  \citet{Bovy2017} for G and K dwarfs.

The measured value of $\alpha$ is interesting from a theoretical perspective because it is inconsistent with the value $\alpha=2$ predicted for an isothermal distribution. \citet{Banerjee} have argued that a steeper value would be expected as a consequence of the constraining effect of the mass in the thin gaseous disk.

The measured scale height $z_e=259$\,pc for the $\sech^\alpha$ profile, or $z_e=269$\,pc for the exponential profile, is broadly in line with previous measurements for the thin disk. A useful comparison is against the result of  \citet{1Bochanski} who measured a scale height of $300\pm15$\,pc from a large sample of early and mid M dwarfs, accounting for binaries and Malmquist bias. This is satisfactory agreement given the much larger distances sampled by their survey, the different analysis method, and the uncertainty in the photometric parallaxes used by them.

The measured height of the Sun above the plane of $10.9^{+1.7}_{-1.6}$\,pc also deserves comment. There have been several measurements of this quantity. One of the most recent and most detailed is the study of \citet{Bennett2019} who find a height $20.8\pm0.3$\,pc. They emphasise the influence of asymmetries in the vertical density distribution, and they define the Galactic plane as the centre of the symmetric part of the density profile. The main asymmetry manifests itself at heights of 500\,pc, beyond the limit of our survey. Since our data look symmetric we might expect the two results to agree somewhat better. Nevertheless as shown in Sec. \ref{sec:binned}, a difference at this level has only a small effect on the estimate of $\alpha$.

\section{The luminosity function}\label{sec:lf}
The $\sech^\alpha$ fits produce an estimated space density at the Galactic plane for each spectral type, Table \ref{tab:results}. We can now compare against measurements of the local space density, in particular the measurements by \citet{Bardalez2019} who have made a comprehensive census of ultracool dwarfs at distances $<25$\,pc. Referring to Fig. \ref{fig:bestfits}, with the Sun located 10\,pc above the plane, the space density in the local bubble of radius 25\,pc will be almost identical to the value at the mid plane. 

\begin{figure}
\centering
\includegraphics[width = 9.cm, trim = 1cm 8cm 0cm 4cm]{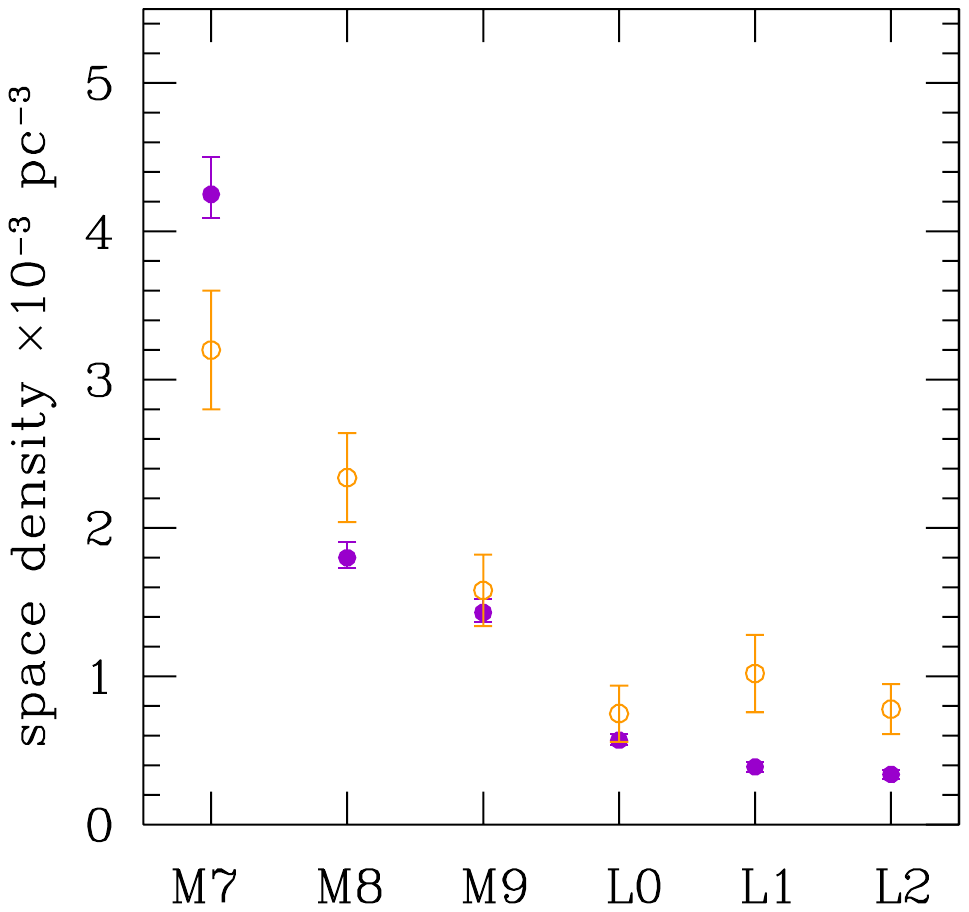}
\caption{Comparison of measured local space density by spectral type. Solid points: this paper. Open points: \citet{Bardalez2019}.}
\label{fig:gagliufi1}
\end{figure}

\begin{table}
\centering
\caption{Comparison of space densities $\mathrm{pc}^{-3}$ in this paper against \citet{Bardalez2019}}
\begin{tabular}{ccc}
\hline\hline
spectral type & this paper & \citet{Bardalez2019}  \\ \hline
M7   & $ 4.25^{+0.25}_{-0.16}\times10^{-3} $ & $ 3.20^{+0.40}_{-0.40}\times10^{-3} $ \\
M8   & $ 1.80^{+0.11}_{-0.07}\times10^{-3} $ & $ 2.34^{+0.30}_{-0.30}\times10^{-3} $ \\
M9   & $ 1.43^{+0.09}_{-0.06}\times10^{-3} $ & $ 1.58^{+0.24}_{-0.24}\times10^{-3} $ \\
L0   & $ 0.57^{+0.04}_{-0.03}\times10^{-3} $ & $ 0.75^{+0.19}_{-0.19}\times10^{-3} $ \\
L1   & $ 0.39^{+0.03}_{-0.03}\times10^{-3} $ & $ 1.02^{+0.26}_{-0.26}\times10^{-3} $ \\
L2   & $ 0.34^{+0.03}_{-0.03}\times10^{-3} $ & $ 0.78^{+0.17}_{-0.17}\times10^{-3} $ \\ \hline
\end{tabular}
\label{tab:resultsphi}
\end{table}

We compare our measurements of the local space density with those of \citet{Bardalez2019} in Table \ref{tab:resultsphi} and Fig. \ref{fig:gagliufi1}. Because \citet{Bardalez2019} use a full spectral sub-type, for e.g. M7 we have combined our results for M7 and M7.5.\footnote{This could in principle lead to small differences. A full sub-type bin effectively covers M6.5 to M7.5. Adding two half sub-type bins effectively covers M6.75 to M7.75.} There is mostly fair agreement between the two determinations, although the points for M7, L1, and L2 are not in statistical agreement (outside $2\sigma$). For the M7 to M9 dwarfs, recall (Sect. \ref{sec:samples}) that \citet{Laithwaite2020} noted an apparent discrepancy between spectral classifications in the homogeneous BOSS sample of \citet{Schmidt2015} and classifications collected from older literature. The difference is in the sense that older measurements found earlier spectral types than measured by \citet{Schmidt2015}. If the classifications in \citet{Bardalez2019} are systematically different to the BOSS classifications for M7 to M9 this would translate to differences in their measured space densities compared to ours.
 
The differences in space density at L1 and L2, where we measure values a factor two smaller than \citet{Bardalez2019}, are harder to understand. The uncertainties plotted on our sample are very small compared to theirs.

\begin{figure}
\centering
\includegraphics[width = 9.cm, trim = 1cm 7cm 0cm 4cm]{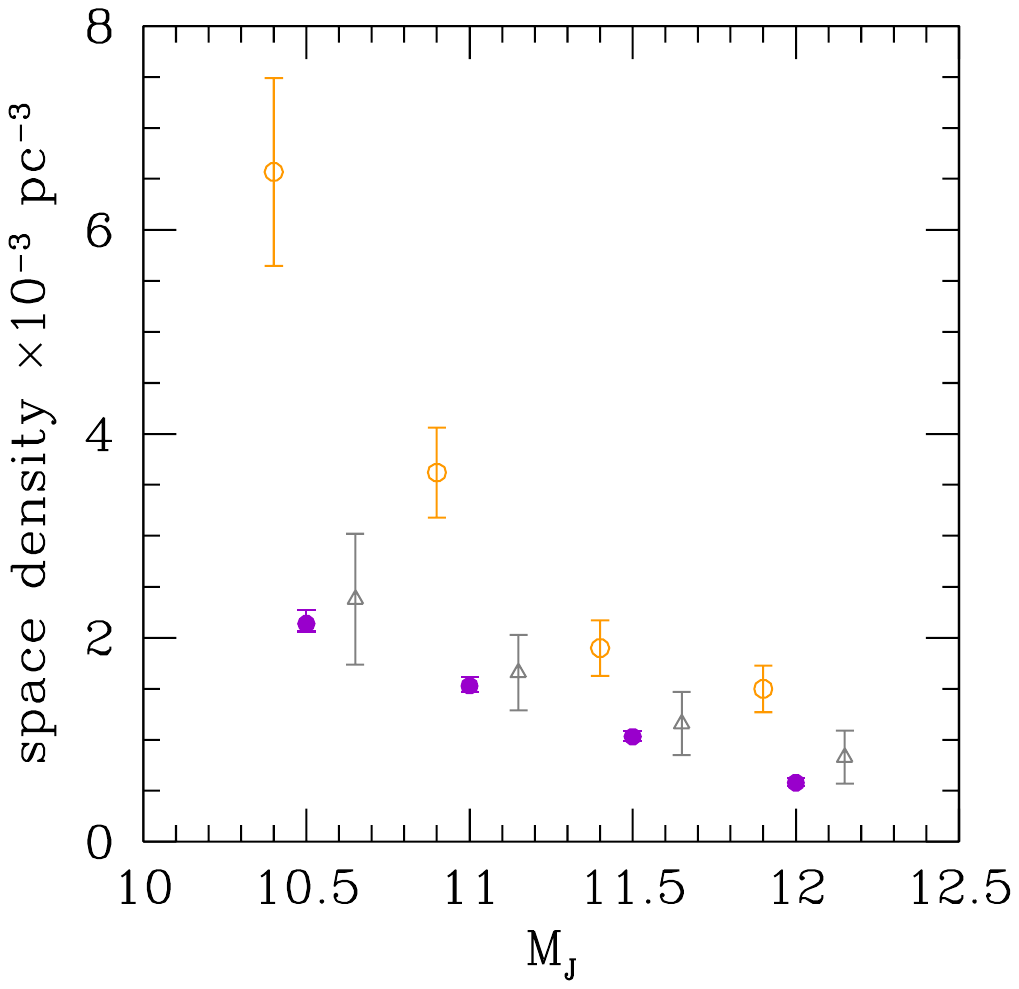}
\caption{Comparison of measured luminosity functions in 0.5\,mag. bins. Solid circles: this paper. Open circles: \citet{Bardalez2019}. Open triangles: \citet{Cruz2007}. The Bardalez Gagliuffi and Cruz points have been shifted 0.1 mag. to the left because they use the 2MASS $J$ band whereas we use the MKO $J$ band.}
\label{fig:gagliufi2}
\end{figure}

\begin{table}
\centering
\caption{Comparison of the luminosity function $\mathrm{pc}^{-3}$ from this paper against the same from \citet{Bardalez2019}}
\begin{tabular}{rcc}
\hline\hline
\multicolumn{1}{c}{$M_J$} & this paper & \citet{Bardalez2019}  \\ 
         & $J_{\mathrm MKO}$  &  $J_{\mathrm 2MASS}$ \\ \hline
$10.25-10.75$   & $ 2.14^{+0.13}_{-0.08}\times10^{-3} $ & $ 6.57^{+0.92}_{-0.92}\times10^{-3} $ \\
$10.75-11.25$   & $ 1.53^{+0.09}_{-0.06}\times10^{-3} $ & $ 3.62^{+0.44}_{-0.44}\times10^{-3} $ \\
$11.25-11.75$   & $ 1.03^{+0.06}_{-0.04}\times10^{-3} $ & $ 1.90^{+0.27}_{-0.27}\times10^{-3} $ \\
$11.75-12.25$   & $ 0.58^{+0.04}_{-0.03}\times10^{-3} $ & $ 1.50^{+0.23}_{-0.23}\times10^{-3} $ \\ \hline
\end{tabular}
\label{tab:resultslf}
\end{table}

Any systematic differences in the spectral classifications between samples would not necessarily translate into differences in the measured luminosity function, as long as absolute magnitudes have been determined correctly for each sample (this would not be the case if the same relation between spectral type and absolute magnitude were used for both samples). To calculate the luminosity function we consider the bins of size 0.5\,mag listed in Table \ref{tab:resultslf} and compute the space density in each bin.\footnote{These have been chosen to match the bins in \citet{Bardalez2019}. We have been informed that where they list a bin as 10.25, this means the range $10.25-10.75$.}  It is important to allow correctly for the spread in absolute magnitude of each spectral sub-type, to include all sub-types that contribute to a bin (given the spread), and to ignore bins where sub-types not considered here would contribute significantly. The last is true for example for the bin $9.75-10.25$, where M6.5 stars would contribute. We assume the absolute magnitudes of each of the twelve spectral sub-types are centred on the values listed in Table \ref{tab:counts} and are Gaussian distributed with $\sigma_M=0.24$ (Sect. \ref{sec:method}). We then integrate the space densities in the MCMC chains into the relevant absolute magnitude bins to compute the luminosity function and uncertainties. For the  highest luminosity bin $10.25-10.75$ any contribution from spectral type M6.5 will be negligible. For the lowest luminosity bin $11.75-12.25$ we have added in a small contribution from L3 dwarfs of 0.02\,$\mathrm{pc}^{-3}$, estimated by assuming the space density of L3 dwarfs is the same as that of L2.5 dwarfs.

The results are listed in Table \ref{tab:resultslf}, where they are compared against the luminosity function results of \citet{Bardalez2019}.  Again we find that the uncertainties quoted by \citet{Bardalez2019} are smaller than a fractional uncertainty of $1/\sqrt{N}$, and so the values quoted in Table  \ref{tab:resultslf} use $1/\sqrt{N}$. Our values for the luminosity function are a factor of two to three lower than those of \citet{Bardalez2019}. Substantially lower space densities in each bin could have been anticipated, since the estimated space densities are lower for most spectral types, Fig. \ref{fig:gagliufi1}, and in addition the range of absolute magnitudes for what we call M7 to M9 is larger than their range, so the space density per 0.5\,mag. bin is further lowered for our sample. The two estimates of the luminosity function are plotted in Fig. \ref{fig:gagliufi2}. Our results use the MKO $J$ band whereas they use the 2MASS $J$ band. As noted earlier, over this spectral range a star has $J_{\mathrm 2MASS}-J_{\mathrm MKO}\sim0.1$\,mag. \citep{Stephens2004}. Therefore we have shifted their points 0.1\,mag to the left in Fig. \ref{fig:gagliufi2} to represent their results on the MKO system. 

In Fig. \ref{fig:gagliufi1} we also plot the older results on the luminosity function from \citet{Cruz2007} (their Table 11) which the \citet{Bardalez2019} results supersede. These also use the 2MASS $J$ band and so the points have also been offset to the left by 0.1 mag (their locations are plotted incorrectly in \citet{Bardalez2019}, offset by 0.25 mag.). Interestingly our results and those of \citet{Cruz2007} agree well.

\section{Summary}\label{sec:summary}

The main points in this paper are the following:

\begin{enumerate}

\item We have used a homogeneous sample of 34\,000 ultracool dwarfs of spectral type M7 to L2.5, all at distances $<350$\,pc, to measure the local vertical density distribution of stars in the disk of the Milky Way. The sample was selected in the $J$ band and benefits from high photometric precision and low extinction. We have developed a likelihood analysis that uses all the stars in the sample optimally, accounting directly for the proportion of unresolved binaries in the sample, and treating Malmquist bias.
\item Fitting the function $\sech^\alpha$ to the density distribution as a function of height from the Galatic plane, we measure $\alpha=0.29^{+0.12}_{-0.13}$. The exponential profile $\alpha=0$ is contained within the $95\%$ credible interval. Any softening of the density distribution towards the plane relative to an exponential profile is modest. The flatter $\sech$ and $\sech^2$ profiles are ruled out at high confidence.
\item Because of the good sampling of the peak of the density distribution the sample is useful for measuring the location of the Galactic plane for this population, and we find the Sun lies at a height $10.9^{+1.7}_{-1.6}$\,pc above the plane.
\item We have used the results of the fit of the density profile to measure the stellar luminosity function at the bottom of the main sequence over the absolute magnitude interval $10.25<M_J<12.25$. Our results for the luminosity function are a factor two to three lower than the measurements by \citet{Bardalez2019} that uses stars in the local 25\,pc radius bubble, but agree well with the older study of \citet{Cruz2007}.

\end{enumerate}

\section*{Acknowledgements}
We are grateful to Daniel Mortlock for a number of helpful discussions on data analysis, and to Daniella Bardalez Gagliuffi for comments on an earlier version of the manuscript. We thank the anonymous referee for comments that led to a number of improvements in the presentation.

\begin{appendices}

\section{Systematics from spectroscopic parallaxes}\label{sec:appendix}

In this appendix we quantify possible systematic errors in the analysis of the vertical density profile due to the adoption of the absolute magnitudes listed in Table \ref{tab:counts}, for each spectral type. The fitting of the vertical density profile assumes that, for example, a M7 star has an absolute magnitude of $M_J=9.92\pm0.24$. For the spectral types M7 to M9.5 the values of $M_J$ were taken from \citet{Laithwaite2020}. They were determined using a volume complete sample of stars with GAIA DR2 parallaxes, where the upper and lower distance limits were a function $G-J$ colour. These distance limits ensured that the sample was representative of the multiplicity of the population. Then the relation between $M_J$ (for single stars) and $G-J$ was determined by fitting the mode of the distribution, assuming a Gaussian distribution in $M_J$ at fixed colour, and a population of equal-mass binaries offset by 0.75\,mag., with the same dispersion. Finally $M_J$ for each spectral type was selected as the value on this relation corresponding to the median colour for that spectral type, and the dispersion of $M_J$ for the population was also measured. The question then arises of how well this model represents the population or whether it could lead to systematic errors in distances because it is too simple.

We can answer this question directly using the original dataset from which the absolute magnitudes were measured. Specifically there are 1737 sources classified M7 in the volume-complete sample of \citet{Laithwaite2020}. Using the double Gaussian model of singles and equal-mass binaries we can assign a probability any source is single or binary, and we select as singles the 1498 sources for which $p(single)>0.5$. Each source has an accurate parallax, with typical parallax/error of 30. We characterise the (true) distribution of distances of this sample by the mean, dispersion, and skewness, measuring respectively $\mu=123.8$\,pc, $\sigma=23.9$\,pc, and $\gamma_1=-0.15$. Now we ask how accurately do the spectroscopic parallaxes represent this distribution. Estimating the distances based on the apparent magnitude $J$ and the adopted $M_J$, we measure $\mu=126.0$\,pc, $\sigma=28.5$\,pc, and $\gamma_1=0.05$. The mean distance estimated using spectroscopic parallaxes is $1.7\%$ larger than the correct value. This fractional difference is significantly smaller than the random uncertainty on the measured disk scale height (Table \ref{tab:results}) and so may be disregarded.

The skewness is very small in both measurements. The standard deviation is larger for the estimated distances compared to the true distances. This is expected, because the estimated distances have a substantial uncertainty of $11\%$. We can check this by convolving the distribution of true distances with this Gaussian error distribution, and then we measure $\sigma=27.9$\,pc, in very good agreement. We conclude that the adopted value of $M_J$ and the Gaussian error distribution provide an accurate model for the distances of the population. The uncertainty in the distances smears out the true distribution of distances, but this smearing is fully accounted for in the fitting, and corrected for, by the procedure used to correct for Malmquist bias, explained in Sect. \ref{sec:method}.

\end{appendices}

\bibliographystyle{ECA_jasa}
\bibliography{paper2ojav2}

@ARTICLE{Cifuentes,
       author = {{Cifuentes}, C. and {Caballero}, J.~A. and {Cort{\'e}s-Contreras}, M. and {Montes}, D. and {Abell{\'a}n}, F.~J. and {Dorda}, R. and {Holgado}, G. and {Zapatero Osorio}, M.~R. and {Morales}, J.~C. and {Amado}, P.~J. and {Passegger}, V.~M. and {Quirrenbach}, A. and {Reiners}, A. and {Ribas}, I. and {Sanz-Forcada}, J. and {Schweitzer}, A. and {Seifert}, W. and {Solano}, E.},
        title = "{CARMENES input catalogue of M dwarfs. V. Luminosities, colours, and spectral energy distributions}",
      journal = {\aap},
         year = 2020,
        month = oct,
       volume = {642},
          eid = {A115},
        pages = {A115},
          doi = {10.1051/0004-6361/202038295},
archivePrefix = {arXiv},
       eprint = {2007.15077},
 primaryClass = {astro-ph.SR},
       adsurl = {https://ui.adsabs.harvard.edu/abs/2020A&A...642A.115C}
}

@ARTICLE{Rix93,
       author = {{Rix}, Hans-Walter and {Rieke}, Marcia J.},
        title = "{Tracing the Stellar Mass in M51}",
      journal = {\apj},
         year = 1993,
        month = nov,
       volume = {418},
        pages = {123},
          doi = {10.1086/173376},
       adsurl = {https://ui.adsabs.harvard.edu/abs/1993ApJ...418..123R}
}

@ARTICLE{Kirkpatrick1999,
       author = {{Kirkpatrick}, J. Davy and {Reid}, I. Neill and {Liebert}, James and
         {Cutri}, Roc M. and {Nelson}, Brant and {Beichman}, Charles A. and
         {Dahn}, Conard C. and {Monet}, David G. and {Gizis}, John E. and
         {Skrutskie}, Michael F.},
        title = "{Dwarfs Cooler than ``M``: The Definition of Spectral Type ``L'' Using Discoveries from the 2 Micron All-Sky Survey (2MASS)}",
      journal = {\apj},
         year = 1999,
        month = jul,
       volume = {519},
       number = {2},
        pages = {802-833},
          doi = {10.1086/307414},
       adsurl = {https://ui.adsabs.harvard.edu/abs/1999ApJ...519..802K}
}

@ARTICLE{Dieterich2014,
       author = {{Dieterich}, Sergio B. and {Henry}, Todd J. and {Jao}, Wei-Chun and
         {Winters}, Jennifer G. and {Hosey}, Altonio D. and {Riedel}, Adric R. and
         {Subasavage}, John P.},
        title = "{The Solar Neighborhood. XXXII. The Hydrogen Burning Limit}",
      journal = {\aj},
         year = 2014,
        month = may,
       volume = {147},
       number = {5},
          eid = {94},
        pages = {94},
          doi = {10.1088/0004-6256/147/5/94},
archivePrefix = {arXiv},
       eprint = {1312.1736},
 primaryClass = {astro-ph.SR},
       adsurl = {https://ui.adsabs.harvard.edu/abs/2014AJ....147...94D}
}

@ARTICLE{Cruz2007,
       author = {{Cruz}, Kelle L. and {Reid}, I. Neill and {Kirkpatrick}, J. Davy and
         {Burgasser}, Adam J. and {Liebert}, James and {Solomon}, Adam R. and
         {Schmidt}, Sarah J. and {Allen}, Peter R. and {Hawley}, Suzanne L. and
         {Covey}, Kevin R.},
        title = "{Meeting the Cool Neighbors. IX. The Luminosity Function of M7-L8 Ultracool Dwarfs in the Field}",
      journal = {\aj},
         year = 2007,
        month = feb,
       volume = {133},
       number = {2},
        pages = {439-467},
          doi = {10.1086/510132},
archivePrefix = {arXiv},
       eprint = {astro-ph/0609648},
 primaryClass = {astro-ph},
       adsurl = {https://ui.adsabs.harvard.edu/abs/2007AJ....133..439C}
}

@ARTICLE{Stephens2004,
       author = {{Stephens}, D.~C. and {Leggett}, S.~K.},
        title = "{JHK Magnitudes for L and T Dwarfs and Infrared Photometric Systems}",
      journal = {\pasp},
         year = 2004,
        month = jan,
       volume = {116},
       number = {815},
        pages = {9-21},
          doi = {10.1086/381135},
archivePrefix = {arXiv},
       eprint = {astro-ph/0311229},
 primaryClass = {astro-ph},
       adsurl = {https://ui.adsabs.harvard.edu/abs/2004PASP..116....9S}
}

@ARTICLE{Tokunaga2002,
       author = {{Tokunaga}, A.~T. and {Simons}, D.~A. and {Vacca}, W.~D.},
        title = "{The Mauna Kea Observatories Near-Infrared Filter Set. II. Specifications for a New JHKL'M' Filter Set for Infrared Astronomy}",
      journal = {\pasp},
         year = 2002,
        month = feb,
       volume = {114},
       number = {792},
        pages = {180-186},
          doi = {10.1086/338545},
archivePrefix = {arXiv},
       eprint = {astro-ph/0110593},
 primaryClass = {astro-ph},
       adsurl = {https://ui.adsabs.harvard.edu/abs/2002PASP..114..180T}
}

@ARTICLE{Camm1950,
       author = {{Camm}, G.~L.},
        title = "{Self-gravitating star systems}",
      journal = {\mnras},
         year = 1950,
        month = jan,
       volume = {110},
        pages = {305},
          doi = {10.1093/mnras/110.4.305},
       adsurl = {https://ui.adsabs.harvard.edu/abs/1950MNRAS.110..305C}
}

@ARTICLE{Lewis2019,
       author = {{Lewis}, Antony},
        title = "{GetDist: a Python package for analysing Monte Carlo samples}",
      journal = {arXiv e-prints},
         year = 2019,
        month = oct,
          eid = {arXiv:1910.13970},
        pages = {arXiv:1910.13970},
archivePrefix = {arXiv},
       eprint = {1910.13970},
 primaryClass = {astro-ph.IM},
       adsurl = {https://ui.adsabs.harvard.edu/abs/2019arXiv191013970L}
}

@ARTICLE{Foreman2013,
       author = {{Foreman-Mackey}, Daniel and {Hogg}, David W. and {Lang}, Dustin and
         {Goodman}, Jonathan},
        title = "{emcee: The MCMC Hammer}",
      journal = {\pasp},
         year = 2013,
        month = mar,
       volume = {125},
       number = {925},
        pages = {306},
          doi = {10.1086/670067},
archivePrefix = {arXiv},
       eprint = {1202.3665},
 primaryClass = {astro-ph.IM},
       adsurl = {https://ui.adsabs.harvard.edu/abs/2013PASP..125..306F}
}

@ARTICLE{Bardalez2019,
       author = {{Bardalez Gagliuffi}, Daniella C. and {Burgasser}, Adam J. and
         {Schmidt}, Sarah J. and {Theissen}, Christopher and
         {Gagn{\'e}}, Jonathan and {Gillon}, Michael and {Sahlmann}, Johannes and
         {Faherty}, Jacqueline K. and {Gelino}, Christopher and
         {Cruz}, Kelle L. and {Skrzypek}, Nathalie and {Looper}, Dagny},
        title = "{The Ultracool SpeXtroscopic Survey. I. Volume-limited Spectroscopic Sample and Luminosity Function of M7-L5 Ultracool Dwarfs}",
      journal = {\apj},
         year = 2019,
        month = oct,
       volume = {883},
       number = {2},
          eid = {205},
        pages = {205},
          doi = {10.3847/1538-4357/ab253d},
archivePrefix = {arXiv},
       eprint = {1906.04166},
 primaryClass = {astro-ph.SR},
       adsurl = {https://ui.adsabs.harvard.edu/abs/2019ApJ...883..205B}
}

@ARTICLE{Marshall1983,
       author = {{Marshall}, H.~L. and {Tananbaum}, H. and {Avni}, Y. and {Zamorani}, G.},
        title = "{Analysis of complete quasar samples to obtain parameters of luminosity and evolution functions}",
      journal = {\apj},
         year = 1983,
        month = jun,
       volume = {269},
        pages = {35-41},
          doi = {10.1086/161016},
       adsurl = {https://ui.adsabs.harvard.edu/abs/1983ApJ...269...35M}
}

@ARTICLE{Schmidt2015,
       author = {{Schmidt}, Sarah J. and {Hawley}, Suzanne L. and {West}, Andrew A. and
         {Bochanski}, John J. and {Davenport}, James R.~A. and {Ge}, Jian and
         {Schneider}, Donald P.},
        title = "{BOSS Ultracool Dwarfs. I. Colors and Magnetic Activity of M and L Dwarfs}",
      journal = {\aj},
         year = "2015",
        month = "May",
       volume = {149},
       number = {5},
          eid = {158},
        pages = {158},
          doi = {10.1088/0004-6256/149/5/158},
archivePrefix = {arXiv},
       eprint = {1410.0014},
 primaryClass = {astro-ph.SR},
       adsurl = {https://ui.adsabs.harvard.edu/abs/2015AJ....149..158S}
}

@ARTICLE{Dupuy2012,
       author = {{Dupuy}, Trent J. and {Liu}, Michael C.},
        title = "{The Hawaii Infrared Parallax Program. I. Ultracool Binaries and the L/T Transition}",
      journal = {\apjs},
         year = 2012,
        month = aug,
       volume = {201},
       number = {2},
          eid = {19},
        pages = {19},
          doi = {10.1088/0067-0049/201/2/19},
archivePrefix = {arXiv},
       eprint = {1201.2465},
 primaryClass = {astro-ph.SR},
       adsurl = {https://ui.adsabs.harvard.edu/abs/2012ApJS..201...19D}
}

@ARTICLE{Reid2008,
       author = {{Reid}, I. Neill and {Cruz}, K.~L. and {Burgasser}, Adam J. and
         {Liu}, Michael C.},
        title = "{L-Dwarf Binaries in the 20-Parsec Sample}",
      journal = {\aj},
         year = 2008,
        month = feb,
       volume = {135},
       number = {2},
        pages = {580-587},
          doi = {10.1088/0004-6256/135/2/580},
       adsurl = {https://ui.adsabs.harvard.edu/abs/2008AJ....135..580R}
}

@ARTICLE{Laithwaite2020,
       author = {{Laithwaite}, R.~C. and {Warren}, S.~J.},
        title = "{The absolute magnitudes $M_J$, the binary fraction, and the binary mass ratios of M7 to M9.5 dwarfs}",
      journal = {arXiv e-prints},
         year = 2020,
        month = jun,
          eid = {arXiv:2006.11092},
        pages = {arXiv:2006.11092},
archivePrefix = {arXiv},
       eprint = {2006.11092},
 primaryClass = {astro-ph.SR},
       adsurl = {https://ui.adsabs.harvard.edu/abs/2020arXiv200611092L}
}

@ARTICLE{Green2018,
       author = {{Green}, Gregory M. and {Schlafly}, Edward F. and {Finkbeiner}, Douglas and
         {Rix}, Hans-Walter and {Martin}, Nicolas and {Burgett}, William and
         {Draper}, Peter W. and {Flewelling}, Heather and {Hodapp}, Klaus and
         {Kaiser}, Nicholas and {Kudritzki}, Rolf-Peter and
         {Magnier}, Eugene A. and {Metcalfe}, Nigel and {Tonry}, John L. and
         {Wainscoat}, Richard and {Waters}, Christopher},
        title = "{Galactic reddening in 3D from stellar photometry - an improved map}",
      journal = {\mnras},
         year = 2018,
        month = jul,
       volume = {478},
       number = {1},
        pages = {651-666},
          doi = {10.1093/mnras/sty1008},
archivePrefix = {arXiv},
       eprint = {1801.03555},
 primaryClass = {astro-ph.GA},
       adsurl = {https://ui.adsabs.harvard.edu/abs/2018MNRAS.478..651G}
}

@ARTICLE{Skrzypek2015,
       author = {{Skrzypek}, N. and {Warren}, S.~J. and {Faherty}, J.~K. and
         {Mortlock}, D.~J. and {Burgasser}, A.~J. and {Hewett}, P.~C.},
        title = "{Photometric brown-dwarf classification. I. A method to identify and accurately classify large samples of brown dwarfs without spectroscopy}",
      journal = {\aap},
         year = 2015,
        month = feb,
       volume = {574},
          eid = {A78},
        pages = {A78},
          doi = {10.1051/0004-6361/201424570},
archivePrefix = {arXiv},
       eprint = {1411.7578},
 primaryClass = {astro-ph.IM},
       adsurl = {https://ui.adsabs.harvard.edu/abs/2015A&A...574A..78S}
}

@ARTICLE{Gould1996,
       author = {{Gould}, Andrew and {Bahcall}, John N. and {Flynn}, Chris},
        title = "{Disk M Dwarf Luminosity Function from Hubble Space Telescope Star Counts}",
      journal = {\apj},
         year = 1996,
        month = jul,
       volume = {465},
        pages = {759},
          doi = {10.1086/177460},
archivePrefix = {arXiv},
       eprint = {astro-ph/9505087},
 primaryClass = {astro-ph},
       adsurl = {https://ui.adsabs.harvard.edu/abs/1996ApJ...465..759G}
}

@ARTICLE{Skrzypek2016,
       author = {{Skrzypek}, N. and {Warren}, S.~J. and {Faherty}, J.~K.},
        title = "{Photometric brown-dwarf classification. II. A homogeneous sample of 1361 L and T dwarfs brighter than J = 17.5 with accurate spectral types}",
      journal = {\aap},
         year = 2016,
        month = may,
       volume = {589},
          eid = {A49},
        pages = {A49},
          doi = {10.1051/0004-6361/201527359},
archivePrefix = {arXiv},
       eprint = {1602.08582},
 primaryClass = {astro-ph.IM},
       adsurl = {https://ui.adsabs.harvard.edu/abs/2016A&A...589A..49S}
}

@ARTICLE{vdk81,
       author = {{van der Kruit}, P.~C. and {Searle}, L.},
        title = "{Surface photometry of edge-on spiral galaxies. I - A model for the three-dimensional distribution of light in galactic disks.}",
      journal = {\aap},
         year = 1981,
        month = feb,
       volume = {95},
        pages = {105-115},
       adsurl = {https://ui.adsabs.harvard.edu/abs/1981A&A....95..105V}
}

@ARTICLE{Dobbie,
       author = {{Dobbie}, Phillip S. and {Warren}, Stephen J.},
        title = "{A Bayesian Approach to the Vertical Structure of the Disk of the Milky Way}",
      journal = {The Open Journal of Astrophysics},
         year = 2020,
        month = jun,
       volume = {3},
       number = {1},
          eid = {5},
        pages = {5},
          doi = {10.21105/astro.2003.05757},
archivePrefix = {arXiv},
       eprint = {2003.05757},
 primaryClass = {astro-ph.GA},
       adsurl = {https://ui.adsabs.harvard.edu/abs/2020OJAp....3E...5D}
}

@ARTICLE{Covey08,
       author = {{Covey}, Kevin R. and {Hawley}, Suzanne L. and {Bochanski}, John J. and
         {West}, Andrew A. and {Reid}, I. Neill and {Golimowski}, David A. and
         {Davenport}, James R.~A. and {Henry}, Todd and {Uomoto}, Alan and
         {Holtzman}, Jon A.},
        title = "{The Luminosity and Mass Functions of Low-Mass Stars in the Galactic Disk. I. The Calibration Region}",
      journal = {\aj},
         year = "2008",
        month = "Nov",
       volume = {136},
       number = {5},
        pages = {1778-1798},
          doi = {10.1088/0004-6256/136/5/1778},
archivePrefix = {arXiv},
       eprint = {0807.2452},
 primaryClass = {astro-ph},
       adsurl = {https://ui.adsabs.harvard.edu/abs/2008AJ....136.1778C}
}

@ARTICLE{Bennett2019,
       author = {{Bennett}, Morgan and {Bovy}, Jo},
        title = "{Vertical waves in the solar neighbourhood in Gaia DR2}",
      journal = {\mnras},
         year = "2019",
        month = "Jan",
       volume = {482},
       number = {1},
        pages = {1417-1425},
          doi = {10.1093/mnras/sty2813},
archivePrefix = {arXiv},
       eprint = {1809.03507},
 primaryClass = {astro-ph.GA},
       adsurl = {https://ui.adsabs.harvard.edu/abs/2019MNRAS.482.1417B}
}

@ARTICLE{Hammersley1999,
       author = {{Hammersley}, P.~L. and {Cohen}, M. and {Garz{\'o}n}, F. and
         {Mahoney}, T. and {L{\'o}pez-Corredoira}, M.},
        title = "{Structure in the first quadrant of the Galaxy: an analysis of TMGS star counts using the SKY model}",
      journal = {\mnras},
         year = "1999",
        month = "Sep",
       volume = {308},
       number = {2},
        pages = {333-363},
          doi = {10.1046/j.1365-8711.1999.02678.x},
archivePrefix = {arXiv},
       eprint = {astro-ph/9906420},
 primaryClass = {astro-ph},
       adsurl = {https://ui.adsabs.harvard.edu/abs/1999MNRAS.308..333H}
}

@ARTICLE{Xiang2018,
       author = {{Xiang}, Maosheng and {Shi}, Jianrong and {Liu}, Xiaowei and
         {Yuan}, Haibo and {Chen}, Bingqiu and {Huang}, Yang and {Wang}, Chun and
         {Wu}, Yaqian and {Tian}, Zhijia and {Huo}, Zhiying and {Zhang}, Huawei and
         {Zhang}, Meng},
        title = "{Stellar Mass Distribution and Star Formation History of the Galactic Disk Revealed by Mono-age Stellar Populations from LAMOST}",
      journal = {\apjs},
         year = "2018",
        month = "Aug",
       volume = {237},
       number = {2},
          eid = {33},
        pages = {33},
          doi = {10.3847/1538-4365/aad237},
archivePrefix = {arXiv},
       eprint = {1807.04592},
 primaryClass = {astro-ph.GA},
       adsurl = {https://ui.adsabs.harvard.edu/abs/2018ApJS..237...33X}
}

@article{1GilmoreReid,
   author = {Gilmore, G and Reid, N},
   title = {New light on faint stars. III - Galactic structure towards the South Pole and the Galactic thick disc},
   journal = {Monthly Notices of the Royal Astronomical Society},
   volume = {202},
   number = {March},
   pages = {1025-1047},
   DOI = {10.1093/mnras/202.4.1025},
   year = {1983},
   type = {Journal Article}
}

@article{1Juric,
   author = {Juri\'c, Mario and Ivezi\'c, Zeljko and Brooks, Alyson and Lupton, Robert H. and Schlegel, David and Finkbeiner, Douglas and Padmanabhan, Nikhil and Bond, Nicholas and Sesar, Branimir and Rockosi, Constance M. and Knapp, Gillian R. and Gunn, James E. and Sumi, Takahiro and Schneider, Donald P. and Barentine, J. C. and Brewington, Howard J. and Brinkmann, J. and Fukugita, Masataka and Harvanek, Michael and Kleinman, S. J. and Krzesinski, Jurek and Long, Dan and Neilsen, Eric H. Jr. and Nitta, Atsuko and Snedden, Stephanie A. and York, Donald G.},
   title = {The Milky Way Tomography with SDSS. I. Stellar Number Density Distribution},
   journal = {The Astrophysical Journal},
   volume = {673},
   number = {2},
   pages = {864-914},
   year = {2008},
   type = {Journal Article}
}

@article{1Bochanski,
   author = {Bochanski, John J. and Hawley, Suzanne L. and Covey, Kevin R. and West, Andrew A. and Reid, I. Neill and Golimowski, David A. and Ivezi\'c, Zeljko},
   title = {The Luminosity and Mass Functions of Low-mass Stars in the Galactic Disk. II. The Field },
   journal = {The Astronomical Journal},
   volume = {139},
   number = {6},
   pages = {2679-2699},
   DOI = {10.1088/0004-6256/139/6/2679},
   year = {2010},
   type = {Journal Article}
}

@article{1Ferguson,
   author = {Ferguson, Deborah and Gardner, Susan and Yanny, Brian},
   title = {Milky Way Tomography with K and M Dwarf Stars: The Vertical Structure of the Galactic Disk},
   journal = {The Astrophysical Journal},
   volume = {843},
   number = {2},
   DOI = {10.3847/1538-4357/aa77fd},
   year = {2017},
   type = {Journal Article}
}

@article{1Chang,
   author = {Chang, Chan-Kao and Ko, Chung-Ming and Peng, Ting-Hung},
   title = {Information on the Milky Way from the Two Micron All Sky Survey Whole Sky Star Count: The Structure Parameters},
   journal = {The Astrophysical J
   ournal},
   volume = {740},
   number = {1},
   DOI = {10.1088/0004-637X/740/1/34},
   year = {2011},
   type = {Journal Article}
}

@article{1Siegel,
   author = {Siegel, M. H. and Majewski, S. R. and Reid, I. N. and Thompson, I. B.},
   title = {Star Counts Redivivus. IV. Density Laws through Photometric Parallaxes},
   journal = {The Astrophysical Journal},
   volume = {578},
   number = {1},
   pages = {151-175},
   DOI = {10.1086/342469},
   year = {2002},
   type = {Journal Article}
}

@article{0Spitzer,
   author = {Spitzer, Lyman},
   title = {The Dynamics of Interstellar Medium III. Galactic Distribution},
   journal = {The Astrophysical Journal},
   volume = {95},
   number = {3},
   pages = {329},
   DOI = {10.1086/144407},
   year = {1942},
   type = {Journal Article}
}

@article{Banerjee,
   author = {Banerjee, Arunima and Jog, Chanda J.},
   title = {The Origin of Steep Vertical Stellar Distribution in the Galactic Disk},
   journal = {The Astrophysical Journal},
   volume = {662},
   pages = {335-340},
   year = {2007},
   type = {Journal Article}
}

@article{deGrijs,
   author = {de Grijs, R. and Peletier, R.F. and van der Kruit, P.C.},
   title = {The z-structure of disk galaxies towards galactic planes},
   journal = {Astronomy and Astrophysics},
   volume = {327},
   pages = {966-982},
   year = {1997},
   type = {Journal Article}
}

@article{Bovy2017,
   author = {Bovy, Jo},
   title = {Stellar Inventory of the Solar Neighborhood using Gaia DR1},
   journal = {Monthly Notices of the Royal Astronomical Society},
   volume = {470},
   pages = {1360-1387},
   year = {2017},
   type = {Journal Article}
}

@article{Kruit1988,
   author = {Van der Kruit, P.C.},
   title = {The three-dimensional distribution of light and mass in disks of spiral galaxies},
   journal = {Astronomy and Astrophysics},
   volume = {192},
   number = {1-2},
   pages = {117-127},
   ISSN = {0004-6361},
   year = {1988},
   type = {Journal Article}
}

@article{Widrow2012,
  title={Galactoseismology: discovery of vertical waves in the Galactic Disk},
  author={Widrow, Lawrence M and Gardner, Susan and Yanny, Brian and Dodelson, Scott and Chen, Hsin-Yu},
  journal={The Astrophysical Journal Letters},
  volume={750},
  number={2},
  pages={L41},
  year={2012},
  publisher={IOP Publishing}
}

@article{Ahmed2019,
  title={A homogeneous sample of 34 000 M7- M9. 5 dwarfs brighter than J= 17.5 with accurate spectral types},
  author={Ahmed, S and Warren, SJ},
  journal={Astronomy \& Astrophysics},
  volume={623},
  pages={A127},
  year={2019},
  publisher={EDP Sciences}
}

\end{document}